\documentclass{article}
\usepackage[english]{babel}
\usepackage[utf8]{inputenc}
\usepackage{color}

\usepackage{graphicx} 
\usepackage{amsmath} 
\usepackage{amssymb} 
\usepackage{amsthm} 
\usepackage{bm} 
\usepackage{tabularx} 
\usepackage{siunitx} 
\usepackage{algorithm} 
\usepackage{algpseudocode} 
\usepackage{caption} 
\usepackage{subcaption} 
\usepackage{fullpage} 
\usepackage{authblk} 

\usepackage{ulem}

\def\eg{{\textit{e.g.},}}
\def\ie{{\textit{i.e.},}}

\newcommand{\REV}[1]{\textcolor{black}{#1}}

\everymath{\displaystyle}
\graphicspath{{img/}}

\title{A higher-order finite element method with unstructured anisotropic mesh adaption for two phase flows with surface tension}
\author{Modesar Shakoor\thanks{Corresponding author} }
\author{Chung Hae Park}

\affil{IMT Lille Douai, Institut Mines-T\'el\'ecom, Univ. Lille, Centre for Materials and Processes, F-59000 Lille, France}

\begin{document}

\maketitle

\begin{abstract}
A novel finite element framework is proposed for the numerical simulation of two phase flows with surface tension. The Level-Set (LS) method with piece-wise quadratic (P2) interpolation for the liquid-gas interface is used in order to reach higher-order convergence rates in regions with smooth interface. A balanced-force implementation of the continuum surface force model is used to take into account the surface tension and to solve static problems as accurately as possible. Given that this requires a balance between the discretization used for the LS function, and that used for the pressure field, an equal-order P2/P2/P2 scheme is proposed for the Navier-Stokes and LS advection equations, which are strongly coupled with each other. This fully implicit formulation is stabilized using the residual-based variational multiscale framework. In order to improve the accuracy and obtain optimal convergence rates with a minimum number of elements, an anisotropic mesh adaption method is proposed where the unstructured mesh is kept as fine as possible close to the zero iso-value of the P2 LS function. Elements are automatically stretched in regions with flat interface in order to keep the complexity fixed during the simulation. The accuracy and efficiency of this approach are demonstrated for two and three dimensional simulations of a rising bubble.\\

\textit{Keywords:}
multiphase flow, level-set, anisotropic mesh, variational multiscale, higher-order, balanced-force
\end{abstract}

\section{Introduction}

Trending industrial applications such as advanced manufacturing (\eg{} liquid composite moulding, additive manufacturing) feature complex multiphase flows where capillary effects have a significant influence at different scales \cite{Nakhoul2018,Mehdikhani2018,Yan2018am}. An increasing research effort has been made in the last decade towards the development of robust and reliable Finite Element (FE) simulation tools for the modeling of multiphase flows with surface tension.\\
First, stabilized FE methods are necessary in order to solve Navier-Stokes equations for a wide range of flow regimes, including turbulent \cite{Brooks1982}. These methods can also be used to satisfy the inf-sup condition when using incompatible velocity-pressure discretizations \cite{Codina2002}. Recent developments based on the Variational MultiScale (VMS) framework are particularly relevant in that regard \cite{Hughes1998}. For instance, the Residual-Based VMS (RBVMS) framework has been proven to be applicable both to reaction-diffusion-advection equations and to the incompressible Navier-Stokes equations in convection-dominated regimes for various choices of discretization \cite{Bazilevs2007,Yan2018iga}.\\
Second, \REV{as shown in Ref. \cite{Shakoor2017}, advanced mesh adaption and motion techniques are required to maintain a conformal FE mesh at interfaces between the different phases of the flow during the simulation. Instead, a front-tracking method is generally employed in order to represent and track these moving and deforming interfaces. Most surface} tension models additionally introduce a mean curvature dependent surface force term at liquid-gas interfaces \cite{Brackbill1992}. The volume of fluid method \cite{Hirt1981}, which tracks the local volume fraction of each phase during the simulation, is hence not sufficient, because it does not capture the morphology of the interface. The interface morphology can be captured by using the Level-Set (LS) method \cite{Osher1988}, which approximates the boundary of each phase through its signed distance function, a.k.a. LS function. Indeed, the normal vector and the mean curvature of an interface are related to the first and second derivatives of its distance function \cite{Sussman1994}. The properties of volume conservation in the volume of fluid method are however lost when switching to the LS method, which has led researchers to consider combinations of those two methods \cite{Sussman2003}. Alternatively, higher-order discretizations using discontinuous Galerkin methods \cite{DiPietro2006} or isogeometric analysis \cite{Yan2018iga}, anisotropic mesh adaption \cite{Bui2012,Nakhoul2018} and semi-Lagrangian particles \cite{Enright2005} have been considered to deal with the same issue, as well as to improve the order of accuracy at the interface.\\
Because \REV{the liquid-gas interface is usually not discretized with a conformal FE mesh}, the implementation of the surface force term is not straightforward. \REV{Both the discontinuity in material properties (\ie{} density and viscosity) and the surface tension term can be integrated accurately by reconstructing the separate liquid and gas domains and the interface at elements intersected by the interface using the extended FE method \cite{Rasthofer2011,Sauerland2011}. This interface reconstruction and integration step can be avoided using} the Continuum Surface Force (CSF) model, where the surface force is transformed into a volume force distributed in a small layer around the interface through the \REV{Dirac} delta function \cite{Brackbill1992}. \REV{Similarly, the discontinuity in material properties can be} distributed in this small layer through a regularized Heaviside function \cite{Sussman1994}. It is important to point out that the regularized Heaviside function depends on the LS function, and that the delta function depends on the first derivative of the regularized Heaviside function, which can be computed analytically \cite{Brackbill1992,Sussman1994}.\\
The regularized transition of the material properties \REV{relies} on the distance property of LS functions \cite{Sussman1994}. This distance property is usually lost during interface motion and deformation. Thus, it should be restored through an operation called LS reinitialization. This can be achieved directly through geometric distance computation \cite{Shakoor2015} or indirectly by solving an Eikonal equation using fast-marching methods \cite{Sethian2000}. Alternatively, indirect LS reinitialization methods based on the solution of nonlinear hyperbolic, parabolic or elliptic equations have been proposed in the literature \cite{Basting2013}. The choice of LS reinitialization method is not obvious and there is no method that has been widely accepted in the literature. Geometric reinitialization is theoretically more robust, because the exact distance to the interface can be computed \cite{Shakoor2015}. However, it is computationally expensive and its generalization to different FE types and orders is not straightforward \cite{Marchandise2007}. Similarly, the generalization of fast-marching methods to unstructured FE meshes requires generalized upwinding criteria \cite{Sethian2000}. Indirect methods based on a nonlinear hyperbolic equation rely on an iterative process which has been shown to be unstable in the presence of interface singularities \cite{Shakoor2015}. Recently, LS reinitialization methods based on nonlinear parabolic or elliptic equations have been developed. Even if their robustness has not been demonstrated in the frame of multiphase flow simulations yet \cite{Basting2013}, these indirect methods are promising as they can easily be implemented for any FE type and order.\\
Spurious non-physical currents near the interface have been reported in the literature, even for static cases \cite{Francois2006,Marchandise2007,Zahedi2012,Lin2018}. \REV{As analyzed in details in Refs. \cite{Zahedi2012,Denner2017}, these currents originate from a mismatch between the pressure discretization and the surface tension model, and numerical errors in the mean curvature calculation. To address the first issue, a balanced-force algorithm can} be employed. \REV{In this algorithm, the} delta function and the normal vector in the CSF model \REV{are} replaced by the numerical gradient of the regularized Heaviside function, and this gradient as well as the pressure gradient \REV{are} projected to the same discretization space \cite{Francois2006}. To avoid those projections, the same discretization \REV{can} be used for the LS function and the pressure. Since the pressure discretization space cannot be larger than that of the velocity, the balanced-force implementation of higher-order LS methods is more difficult.\\
Last, interface motion in the LS method is modeled through advection equations for LS functions. The advection velocity is no other than the flow velocity computed by the flow solver. The flow model includes the CSF model, which depends on the second derivatives of the LS function and, in a balanced-force implementation, on the first derivatives of the LS-dependent regularized Heaviside function. Thus, a choice of time stepping scheme and coupling should be made. To avoid severe restrictions on the time step \cite{Sussman1994}, various implicit schemes have been considered in the literature. It has been shown that both a semi-implicit and a fully implicit implementation of the CSF model can be achieved without strong coupling of the flow and LS advection solvers \cite{Hysing2006}. To eliminate all restrictions on the time step, fully implicit implementations with strong coupling of the two solvers have been proposed \cite{Pochet2013,Yan2018iga}.\\

In this work, we propose an adaptive fully implicit RBVMS method for two phase flow with surface tension relying on a piece-wise quadratic (P2) discretization of the velocity, the pressure and the LS function as well, in order to reduce the computational cost while maintaining a good accuracy of the numerical solutions. All variables, \ie{} the velocity, the pressure and the LS function are solved simultaneously because Navier-Stokes equations with the CSF model are strongly coupled to the LS advection equation, as opposed to most conventional approaches. The use of the same discretization space for all variables is only possible because of RBVMS stabilization, and enables us to obtain a balanced-force implementation in a straightforward manner without the procedure of pressure gradient projection. Moreover, a large time step can be adopted through the fully implicit RBVMS method while avoiding divergence of the solution. In order to improve the accuracy near the interface, the unstructured mesh is dynamically adapted during the simulation based on an anisotropic P2 error estimator.\\
The governing equations for incompressible two phase flow with surface tension are summarized in Sec. \ref{sec:equations}, and the proposed numerical method is presented in Sec. \ref{sec:method}. In Sec. \ref{sec:results}, the performance of the proposed method in terms of accuracy and efficiency is demonstrated for two and three dimensional simulations of a rising bubble. It is shown that our fully implicit implementation leads to a reduction of the number of time steps compared to other methods proposed in the literature, while anisotropic unstructured mesh adaption allows us to achieve a more optimal FE discretization with a higher-order convergence rate.

\section{\label{sec:equations}Governing equations}

\subsection{Navier-Stokes equations}

The computation domain is denoted as $\Omega \subset \mathbb{R}^d, d=2,3$. The Navier-Stokes momentum and mass conservation equations for isothermal, Newtonian, incompressible flow in $\Omega$ are
\begin{equation}\label{eq:navier-stokes}
\left\lbrace\begin{array}{rcl}
    \rho \left(\frac{\partial \mathbf{v}}{\partial t} + \mathbf{v} . \nabla \mathbf{v}\right) \REV{- \nabla . \mathbf{\sigma}(\mathbf{v},p)} - \mathbf{f}_g + \mathbf{f}_{s} &=& \mathbf{0},\\
    \nabla . \mathbf{v} &=& 0.
\end{array}\right.
\end{equation}
In these equations $\mathbf{v}$ is the velocity vector, $p$ the pressure, $t$ the time, $\rho$ the mass density, and $\mu$ the viscosity. The Cauchy stress tensor \REV{$\mathbf{\sigma}$ is defined as
\begin{equation}\label{eq:stress}
\mathbf{\sigma}(\mathbf{v},p) = \mu (\nabla \mathbf{v} + \nabla^T \mathbf{v}) - p\mathbf{I},
\end{equation}
where $\mathbf{I}$ is the identity matrix.} The force $\mathbf{f}_g = -\rho g \mathbf{e}_y$ is the gravity force, with $g$ the gravity acceleration. The force $\mathbf{f}_{s}$ is the surface tension force and is described in Subsec. \ref{subsec:equations:csf}.\\
Those equations will be accompanied by either Dirichlet or homogeneous Neumann boundary conditions in the paper.

\subsection{LS method}

The computation domain is split into two parts $\Omega_1$ and $\Omega_2$, each occupied by a different fluid, with
$$\Omega_1 \cup \Omega_2 = \Omega\text{, and }\Omega_1 \cap \Omega_2 = \Gamma_{1,2}.$$
Both domains as well as their interface evolve in time according to the motion of the two fluids:
$$\Omega_1=\Omega_1(t), \Omega_2=\Omega_2(t)\text{, and }\Gamma_{1,2}=\Gamma_{1,2}(t).$$\\
Instead of using a conformal FE mesh of $\Gamma_{1,2}$, we introduce the LS function $\phi$ \cite{Osher1988} which is the signed distance function to $\Gamma_{1,2}$:
\begin{equation}\label{eq:level-set}
\phi(\mathbf{x},t) = \left\lbrace\begin{array}{ll}
    +dist(\mathbf{x},\Gamma_{1,2}(t)),& \mathbf{x} \in \Omega_1(t),\\
    -dist(\mathbf{x},\Gamma_{1,2}(t)),& \mathbf{x} \in \Omega_2(t),\\
    0,& \mathbf{x} \in \Gamma_{1,2}(t).
\end{array}\right.
\end{equation}

The first (resp. second) fluid hence occupies the part of the domain with positive (resp. negative) $\phi$, while the interface is defined as the zero isolevel of $\phi$. The evolution of the interface can then be modeled by advecting $\phi$ with the flow velocity:
\begin{equation}\label{eq:level-set-advection}
\frac{\partial \phi}{\partial t} + \mathbf{v} . \nabla \phi = 0.
\end{equation}

Fluid properties in Eq. \eqref{eq:navier-stokes} can now be defined in terms of the LS function \cite{Sussman1998}:
\begin{equation}\label{eq:flow-properties}
\begin{array}{rcl}
\rho(\mathbf{x},t)=\rho(\phi(\mathbf{x},t)) &=& H(\phi(\mathbf{x},t))\rho_1 + (1-H(\phi(\mathbf{x},t)))\rho_2\\
\mu(\mathbf{x},t)=\mu(\phi(\mathbf{x},t)) &=& H(\phi(\mathbf{x},t))\mu_1 + (1-H(\phi(\mathbf{x},t)))\mu_2
\end{array}
\end{equation}
where $\rho_1,\rho_2$ and $\mu_1,\mu_2$ are the mass densities and viscosities of each fluid, respectively, and $H$ is the Heaviside function:
\begin{equation}\label{eq:heaviside}
H(\phi) = \left\lbrace\begin{array}{ll}
    1,& \phi > 0,\\
    \frac{1}{2},& \phi = 0,\\
    0,& \phi < 0.
\end{array}\right.
\end{equation}

\subsection{\label{subsec:equations:csf}CSF model}

The surface tension force $\mathbf{f}_{s}$ can be expressed through the CSF model \cite{Brackbill1992} as
\begin{equation}\label{eq:csf}
\mathbf{f}_{s} = \sigma_s \kappa_s \mathbf{n}_s \delta
\end{equation}
where $\sigma_s$ is the surface tension coefficient, $\kappa_s$ is the mean curvature of the interface, $\mathbf{n}_s$ its normal vector, and \REV{$\delta = \frac{\mathrm{d}H}{\mathrm{d}\phi}$ is the so-called interface density. The computation of this derivative requires a regularization of the Heaviside function, as described in Subsec. \ref{subsec:method:csf}.}\\
The normal vector and the mean curvature can be directly computed from LS function $\phi$ using
\begin{equation}\label{eq:normal-curvature}
\mathbf{n}_s = \mathbf{n}_s(\phi) = \frac{\nabla \phi}{|\nabla \phi|}, \kappa_s = \kappa_s(\phi) = -\nabla . \mathbf{n}_s(\phi) = \frac{\nabla \phi . \nabla \nabla \phi . \nabla \phi - |\nabla \phi|^2 \text{tr}(\nabla \nabla \phi)}{|\nabla \phi|^3},
\end{equation}
where $\nabla \nabla \phi$ is the Hessian matrix of $\phi$ and $|.|$ the Euclidean norm. Due to the sign of the LS function in Eq. \eqref{eq:level-set}, the normal vector will be pointing towards phase 1, which means that the liquid phase is phase 1 while the gas phase is phase 2.\\

The numerical formulation described in the following section solves Eqs. \eqref{eq:navier-stokes} and \eqref{eq:level-set-advection} to approximate $\lbrace\mathbf{v},p,\phi\rbrace$. It can be observed that the LS advection equation depends linearly on the flow velocity, while the momentum equation involves a nonlinear dependence of the mass density and the viscosity on the LS function. This is added to the nonlinearity of the auto-advection term and of the surface tension force.

\section{\label{sec:method}Numerical methods}

In this section, we present the different numerical methods to solve the equations described in Sec. \ref{sec:equations}. Those equations are solved using the RBVMS framework where we choose equal-order FE discretizations $\lbrace\mathbf{v}_h,p_h,\phi_h\rbrace$ for the velocity, pressure and LS solutions. This approximation space is denoted by $\mathcal{V}_h$. The parameter $h$ is proportional to the length of the longest edge of the FE mesh.

\subsection{\label{subsec:method:csf}Balanced-force implementation}

As mentioned in the introduction, to avoid dealing explicitly with the discontinuity in Eqs. \eqref{eq:heaviside} and \eqref{eq:csf}, a regularized Heaviside function $H_{\epsilon}$ will be used in practice instead of $H$ in Eq. \eqref{eq:flow-properties}:
\begin{equation}\label{eq:regularized-heaviside}
H_{\epsilon}(\phi) = \left\lbrace\begin{array}{ll}
    1,& \phi > \epsilon,\\
    \frac{1}{2}\left(1 + \frac{\phi}{\epsilon} + \frac{1}{\pi}\sin\left(\frac{\pi \phi}{\epsilon}\right) \right),& |\phi| \leq \epsilon, \\
    0,& \phi < -\epsilon.
\end{array}\right.
\end{equation}
As a result, the transition of fluid properties within a small layer of thickness $2\epsilon$ around the interface can become smooth. The parameter $\epsilon \sim \mathcal{O}(h)$ should be as small as possible to have an accurate transition of material properties while it cannot be smaller than $\frac{h}{2}$ (as compared to $h$ for P1 FEs \cite{Pochet2013}).\\
In general \cite{Sussman1994,Sussman1998,Yan2018iga}, the $\delta$ function in Eq. \eqref{eq:csf} is also replaced by a regularized function $\delta_{\epsilon}$ which is defined by the analytical derivative of $H_{\epsilon}$, namely $\delta_{\epsilon} = \frac{\mathrm{d}H_{\epsilon}}{\mathrm{d}\phi}$. This function may be weighed to deal with large ratios of mass density \cite{Brackbill1992,Marchandise2007}.\\
However, such formulation leads to oscillations in the case of a static bubble or droplet submitted to no gravity \cite{Francois2006}. In this case Eq. \eqref{eq:level-set-advection} has no effect and Eq. \eqref{eq:navier-stokes} reduces to $\nabla p = \sigma_s \kappa_s \mathbf{n}_s \delta$. Regardless of the numerical method used to estimate the mean curvature, if the discretization space of $\nabla p_h$ does not match that of $\mathbf{n}_s(\phi_h) \delta(\phi_h)$, numerical errors will induce spurious non-physical currents \cite{Francois2006}.\\
To avoid these oscillations, we replace $\mathbf{n}_s(\phi_h) \delta(\phi_h)$ by the numerical gradient of the Heaviside function $\nabla H_{\epsilon}(\phi_h)$, which leads to the following balanced-force implementation of the CSF model:
\begin{equation}\label{eq:balanced-csf}
\mathbf{f}_{s}(\phi_h) = \sigma_s \kappa_s(\phi_h) \nabla H_{\epsilon}(\phi_h).
\end{equation}
where $H_{\epsilon}(\phi_h)$ is defined in Eq. \eqref{eq:regularized-heaviside} and $\kappa_s(\phi_h)$ is computed using Eq. \eqref{eq:normal-curvature}. This is a balanced-force implementation because we are using equal-order discretizations for the pressure, the LS function and the regularized Heaviside function.\\
In order to compute $\kappa_s(\phi_h)$ directly through Eq. \eqref{eq:normal-curvature}, $\phi_h$ needs to be at least piece-wise quadratic. \REV{Alternatively, we could use integration by part to eliminate the computation of second derivatives of the LS function \cite{Rasthofer2011,Sauerland2011}. Nevertheless, the choice of a piece-wise quadratic LS function is adopted in this paper to improve mass and energy conservation \cite{Pochet2013}.} As a consequence, $p_h$ also needs to be piece-wise quadratic and, even though RBVMS stabilization is used, $\mathbf{v}_h$ needs to be at least piece-wise quadratic because of the inf-sup condition. If we wanted to use a higher-order discretization for $\phi_h$ and obtain an improved mass conservation \cite{DiPietro2006}, we could avoid those restrictions by projecting $\nabla H_{\epsilon}(\phi_h)$ down to the pressure gradient discretization space. However, this would deteriorate the accuracy of the discrete CSF model. Consequently, we adopt a P2/P2/P2 discretization.

\subsection{\label{subsec:method:rbvms}Fully implicit RBVMS framework}

The stabilization of advection-dominated equations, in particular Navier-Stokes equations in the convection-dominated regime, has been a very active research topic for the last decades \cite{Brooks1982,Hughes1998,Codina2002}. The VMS framework proposed in 1998 has given interesting insights on how this stabilization could be achieved by approximating those equations simultaneously at two scales (\ie{} coarse and fine) and projecting the fine scale onto the coarse scale \cite{Hughes1998}. The coarse one is simply the one discretized by the mesh, while the subgrid scale should be modeled separately. The difficulty resides in how the fine scale is modeled and how it is projected to stabilize the coarse scale. This issue can be addressed through the RBVMS framework which has been proposed for incompressible Navier-Stokes equations \cite{Bazilevs2007} and later on extended to the simulation of multiphase flow with surface tension \cite{Rasthofer2011,Yan2018iga}.\\
In the RBVMS framework for multiphase flow \cite{Yan2018am}, the velocity, pressure and LS approximations are written as
\begin{equation}
\begin{array}{rl}
    \mathbf{v} & \approx \mathbf{v}_h + \mathbf{v}'\\
    p & \approx p_h + p'\\
    \phi & \approx \phi_h + \phi'
\end{array}
\end{equation}
where $\lbrace\mathbf{v}_h,p_h,\phi_h\rbrace$ are the coarse scale solutions and $\lbrace\mathbf{v}',p',\phi'\rbrace$ are the fine scale solutions. The discrete space of velocity, pressure and LS test functions $\lbrace\mathbf{w}_h,q_h,\psi_h\rbrace$ is denoted as $\mathcal{W}_h$. The semi-discrete weak form of Eqs. \eqref{eq:navier-stokes} and \eqref{eq:level-set-advection} follows \cite{Yan2018iga}:\\
Find $\lbrace\mathbf{v}_h,p_h,\phi_h\rbrace \in \mathcal{V}_h^{d+2}$, such that,
\begin{equation}\label{eq:rbvms}
\left\lbrace\begin{array}{ll}
    \int_{\Omega_h} \mathbf{w}_h . \rho(\phi_h)\left(\frac{\partial \mathbf{v}_h}{\partial t} + \mathbf{v}_h . \nabla \mathbf{v}_h\right) \mathrm{d}\Omega&\\
    + \int_{\Omega_h} \nabla \mathbf{w}_h : \mu(\phi_h)\left(\nabla \mathbf{v}_h + \nabla^T \mathbf{v}_h\right) \mathrm{d}\Omega - \int_{\Omega_h} p_h \nabla . \mathbf{w}_h \mathrm{d}\Omega&\\
    - \int_{\Omega_h} \mathbf{w}_h . \mathbf{f}_g(\phi_h) \mathrm{d}\Omega + \int_{\Omega_h} \mathbf{w}_h . \mathbf{f}_s(\phi_h) \mathrm{d}\Omega&\\
    - \sum_{K \in \mathcal{T}_h} \int_{K} \left( \rho(\phi_h) \mathbf{v}_h . \nabla \mathbf{w}_h \right) . \mathbf{v}'(\mathbf{v}_h,p_h,\phi_h) \mathrm{d}\Omega&\\
    + \sum_{K \in \mathcal{T}_h} \int_{K} \rho(\phi_h) \mathbf{w}_h . \left(\nabla \mathbf{v}_h . \mathbf{v}'(\mathbf{v}_h,p_h,\phi_h) \right) \mathrm{d}\Omega&\\
    - \sum_{K \in \mathcal{T}_h} \int_{K} \rho(\phi_h) \nabla \mathbf{w}_h : \left( \mathbf{v}'(\mathbf{v}_h,p_h,\phi_h) \otimes \mathbf{v}'(\mathbf{v}_h,p_h,\phi_h) \right) \mathrm{d}\Omega&\\
    - \sum_{K \in \mathcal{T}_h} \int_{K} \nabla . \mathbf{w}_h p'(\mathbf{v}_h,p_h,\phi_h) \mathrm{d}\Omega& = 0, \forall \mathbf{w}_h \in \mathcal{W}_h^{d},\\
    - \int_{\Omega_h} q_h \nabla . \mathbf{v}_h \mathrm{d}\Omega&\\
    + \sum_{K \in \mathcal{T}_h} \int_{K} \nabla q_h . \mathbf{v}'(\mathbf{v}_h,p_h,\phi_h) \mathrm{d}\Omega& = 0, \forall q_h \in \mathcal{W}_h,\\
    + \int_{\Omega_h} \psi_h \left(\frac{\partial \phi_h}{\partial t} + \mathbf{v}_h . \nabla \phi_h \right) \mathrm{d}\Omega&\\
    - \sum_{K \in \mathcal{T}_h} \int_{K} \nabla \psi_h . \mathbf{v}_h \phi'(\mathbf{v}_h,\phi_h) \mathrm{d}\Omega &= 0, \forall \psi_h \in \mathcal{W}_h,
\end{array}\right.
\end{equation}
where we have indicated explicitly the nonlinear dependence of some terms on the solution. The boundary integral term is omitted due to homogeneous Neumann boundary conditions. The fine scale stabilization terms are integrated element-wise, where $\mathcal{T}_h$ is the FE mesh which is a set of elements spanning the discrete domain $\Omega_h \approx \Omega$. The fine scale stabilization term for the conservation equation makes this formulation inf-sup stable and enables us to use equal-order FEs for velocity and pressure. Readers are referred to Ref. \cite{Bazilevs2007} for the details regarding the remaining fine scale terms. In the RBVMS framework, the fine scale solutions are simply modeled as proportional to the residuals of the strong form coarse scale equations:
\begin{equation}\label{eq:rbvms-fine}
\begin{array}{ll}
    \mathbf{v}'(\mathbf{v}_h,p_h,\phi_h) &= -\frac{\tau_{\mathbf{v}}(\mathbf{v}_h,\phi_h)}{\rho(\phi_h)} \left( \rho(\phi_h) \left(\frac{\partial \mathbf{v}_h}{\partial t} + \mathbf{v}_h . \nabla \mathbf{v}_h\right) - \nabla . \mathbf{\sigma}(\mathbf{v}_h,p_h,\phi_h) - \mathbf{f}_g(\phi_h) + \mathbf{f}_{s}(\phi_h) \right),\\
    p'(\mathbf{v}_h,p_h,\phi_h) &= -\tau_p(\mathbf{v}_h,\phi_h) \rho(\phi_h) \nabla . \mathbf{v}_h,\\
    \phi'(\mathbf{v}_h,\phi_h) &= -\tau_\phi(\mathbf{v}_h) \left(\frac{\partial \phi_h}{\partial t} + \mathbf{v}_h . \nabla \phi_h \right),
\end{array}
\end{equation}
\REV{where $\sigma$ is the Cauchy stress tensor defined in Eq. \eqref{eq:stress}.}
Here we have again indicated explicitly the dependence of the fine scale solutions on the coarse scale solutions as well as that of the following stabilization parameters:
\begin{equation}\label{eq:rbvms-stab}
\begin{array}{ll}
    \tau_{\mathbf{v}}(\mathbf{v}_h,\phi_h) &= \left( \left(\frac{2}{\Delta t}\right)^2 + \mathbf{v}_h . \mathbf{G} \mathbf{v}_h + C_I \left(\frac{\mu(\phi_h)}{\rho(\phi_h)}\right)^2 \mathbf{G} : \mathbf{G} \right)^{-\frac{1}{2}},\\
    \tau_p(\mathbf{v}_h,\phi_h) &= \left( \mathrm{tr}(\mathbf{G})\tau_{\mathbf{v}}(\mathbf{v}_h,\phi_h) \right)^{-1},\\
    \tau_\phi(\mathbf{v}_h) &= \left( \left(\frac{2}{\Delta t}\right)^2 + \mathbf{v}_h . \mathbf{G} \mathbf{v}_h \right)^{-\frac{1}{2}}.
\end{array}
\end{equation}
The constant $C_I$ depends on the type and order of the FE, $\Delta t$ is the time step, and $\mathbf{G}=\mathbf{B}^T\mathbf{B}$ is the element metric tensor, with $\mathbf{B}$ the transposed inverse of the gradient of the mapping from reference element to mesh element \cite{Bazilevs2007}. Note that those stabilization terms are defined and computed at quadrature points within each element, contrary to other stabilization methods where they may be defined as constant element-wise \cite{Brooks1982,Codina2002}.\\
For the time discretization, we use the backward Euler scheme which is unconditionally stable but requires a Newton-Raphson scheme to deal with the nonlinear terms in Eq. \eqref{eq:rbvms}, the fine scale solutions in Eq. \eqref{eq:rbvms-fine} and the stabilization parameters in Eq. \eqref{eq:rbvms-stab}. In order to avoid the difficulty of computing all analytical derivatives, we adopt numerical differentiation. This procedure is conducted locally inside each element by iteratively adding and then removing a small perturbation $\delta_{ND}$ to the node-wise values of each of the solutions $\lbrace\mathbf{v}_h,p_h,\phi_h\rbrace$. The drawback of this approach is that the Newton-Raphson scheme will not converge up to a very low tolerance depending on the value of $\delta_{ND}$.\\

If the prescribed time step $\Delta t_{max}$ is too big or the solution is irregular, the Newton-Raphson scheme might not \REV{converge in a prescribed maximum number of iterations or even} diverge and obtain solutions which lead to a larger residual from one nonlinear iteration to another. If such situation arises, we restart the Newton-Raphson scheme with half the time step and repeat this procedure until we \REV{find a time step small enough for the Newton-Raphson scheme to obtain a converged solution in less than 10 iterations}. Inversely, when a time increment can be solved at once with no restart, we multiply the time step by 1.5, unless it leads to a larger value than $\Delta t_{max}$. \REV{In all simulations presented in this paper, this automatic time step control algorithm was always successful in making the Newton-Raphson scheme converge in less than 10 iterations. As observed in Sec. \ref{sec:results}, this automatic variation of the time step did not lead to an increase of the number of time increments higher than a factor of two.}

\subsection{\label{subsec:method:reinit}LS reinitialization}

\REV{The use of the regularized Heaviside function as defined in Eq. \eqref{eq:regularized-heaviside} requires $\phi$ to remain a signed distance function throughout the simulation.} As mentioned in the introduction, the distance property is lost during interface motion and deformation, and should be restored using LS reinitialization. Because the implicit scheme enables the use of a large prescribed time step $\Delta t_{max}$, we reinitialize $\phi_h$ at every time step.\\
Recently developed indirect methods based on a nonlinear elliptic equation are promising because, in theory, this nonlinear elliptic equation can be easily solved using an FE method with any FE type or order \cite{Basting2013}. Additionally, it should be theoretically possible to integrate this approach directly into the fully implicit RBVMS framework as an additional regularity constraint \cite{Basting2017} and even to add a mass conservation constraint to reduce mass loss during advection and reinitialization \cite{Basting2017}.\\
In the present work, we adopt a more straightforward approach as we opt for direct geometric reinitialization \cite{Shakoor2015}. The implementation of a P2 LS function is technically challenging because the intersection of each element of the mesh with the zero isolevel of $\phi_h$ is not necessarily planar. To avoid this problem, we hierarchically subdivide each P2 FE twice and reconstruct a fine surface mesh of the zero isolevel of $\phi_h$ as a P1 mesh of straight line segments (2D) or flat triangles (3D). More details on this method can be found in Ref. \cite{Marchandise2007}.\\
The distance of any point to the interface $\Gamma_{1,2}$ is approximated by looking for the closest projection among all elements of the surface mesh \cite{Eberly2003}. This search can be conducted efficiently using a space partitioning technique \cite{Bentley1975,Shakoor2015}. In a recent work \cite{Park2010}, it has been shown that Graphics Processing Units (GPUs) can be much more efficient for this kind of task than conventional CPU implementations. Therefore, we opt for a brute force search where we project all mesh nodes on all elements of the surface mesh, with no space partitioning at all. This search is conducted independently for each mesh node on a separate GPU thread. Additionally, we let the LS reinitialization procedure automatically truncate the LS function around the interface so that its values be restricted to $[-2\epsilon,2\epsilon]$.

\subsection{\label{subsec:method:adapt}Anisotropic P2 mesh adaption}

Thanks to the higher-order discretization of the solution $\lbrace\mathbf{v}_h,p_h,\phi_h\rbrace$, it is expected to improve the accuracy and the conservation of the LS function at each time increment compared to P1 methods. The accuracy should be controlled near to the interface, in the region of thickness $2\epsilon$ where we operate the fluid properties transition defined in Eqs. \eqref{eq:regularized-heaviside} and \eqref{eq:flow-properties} and integrate the surface tension force defined in Eq. \eqref{eq:balanced-csf}.\\
The challenge is that the interface is evolving during the simulation. We hence propose to dynamically adapt and refine the unstructured mesh during the simulation to track the interface and keep elements refined in the transition region. To reach this objective, we define a sensor variable $s_h = (H_\epsilon(\phi^{n}_h),H_\epsilon(2\phi^n_h-\phi^{n-1}_h))$, where $\phi^{n}_h$ is the discrete LS function at the start of a time increment (before Newton-Raphson solution) and $2\phi^n_h-\phi^{n-1}_h$ is an extrapolation of $\phi^{n+1}_h$. We do not include the velocity and pressure fields in the sensor variable in order to avoid difficulties in taking into account errors of different natures. This will be investigated in a future work as it could be relevant for multiphysics problems. It can be noted that, because of the use of an equal-order P2/P2/P2 FE method, the convergence rate for the velocity and pressure fields should be improved in the transition region.\\

Isotropic mesh adaption can be based on any scalar \textit{a priori} or \textit{a posteriori} estimation of the approximation error at integration points or mesh nodes. For instance, the well-known Zienkiewicz-Zhu error estimator \cite{Zienkiewicz1987} considers the point-wise difference between the numerical solution and the improved approximation recovered using an operation called Superconvergent Patch Recovery (SPR). This SPR operation which is used subsequently in this work, consists in recovering a higher-order and higher-regularity approximation of the solution around each mesh node. For a current P1 approximation at a given node, the recovered approximation should be of order 2 \cite{Zhang2005}, namely it should embed a recovered node-wise continuous gradient and Hessian matrix. This recovered approximation is fitted in a least-squares sense to the values of the numerical solution at neighboring mesh nodes which is called the patch.\\
Alternatively, the fine scale solutions of the VMS framework described in Subsec. \ref{subsec:method:rbvms} can be used as estimates of the error in the coarse scale approximations \cite{Hughes1998,Bazilevs2007}.\\
The latter interpolation error can be used as a target for error estimation because the approximation error is bounded by it. In particular, the interpolation error for a P1 solution can be estimated from the Hessian matrix of this solution and that for a P2 solution from its third derivatives \cite{Coulaud2016} which can be recovered node-wise using a SPR one order higher than that of the derivatives to be recovered \cite{Belhamadia2004}, namely a SPR of order 3 for second derivatives and order 4 for third derivatives.\\
Elements can be refined and coarsened based on the point-wise estimation of the approximation or interpolation error.\\

For anisotropic mesh adaption, additional directional information on the error is required not only to refine and coarsen elements, but also to stretch them differently along each direction. Improved convergence rates are obtained with unstructured anisotropic mesh adaption, because the mesh can be distorted and oriented locally depending on the estimated error \cite{Belhamadia2004,Coupez2011,Coulaud2016}.\\
In the P1 case, the interpolation error estimator reduces to a Hessian-based error estimator \cite{Alauzet2006,Dobrzynski2008,Bui2012}. A particular feature of this estimator is that the Hessian matrix defines locally a scalar product and a distance measure which are distorted compared to the Euclidean ones \cite{Arsigny2006}. This distortion or stretching depends on the eigenvalues and eigenvectors of the Hessian matrix. The eigenvectors can also be regarded as the axis of an ellipsoid and the square root of the inverse of the eigenvalues as the radii of this ellipsoid \cite{Laug2013}. An edge-based version of the Hessian-based error estimator has been proposed recently and applied to multiphase flow simulations \cite{Coupez2011,Nakhoul2018}.\\
To the best of the authors' knowledge, there has been no anisotropic error estimator based on the fine scale solutions of the VMS framework in the literature except the anisotropic versions of the Zienkiewicz-Zhu estimator \cite{Farrell2011,Porta2012}. Anisotropic \textit{a priori} error estimators for multiphase flow simulations directly targeting the error in the geometric representation of the interface through the LS function have been proposed in Refs. \cite{Compere2008,Xie2016}. In the following, we opt for an error estimator targeting the interpolation error of the sensor variable. This will allow us not only to adapt the mesh during the simulation but also pre-adapt the mesh before solving any equation so as to ensure that the interpolation of the initial LS function is accurate.\\

To prepare an unstructured mesh for the initial time increment or any current time increment, it is required to estimate the interpolation error of $s_h$, and then adapt the mesh to distribute this error uniformly on the domain, for a given complexity. Indeed, we will not try to prescribe an error tolerance, which may blow up the computational cost if the interface area increases during the simulation. Instead, we keep the complexity (the number of P1 nodes of the mesh) constant during the simulation and expect the error estimator to capture only a certain level of detail which can be afforded with the prescribed complexity. Such approaches have been proposed for instance in Refs. \cite{Alauzet2006,Coupez2011} for anisotropic P1 interpolation.\\
Approaches to dynamically adapt an unstructured mesh to control the interpolation error for P1 FEs have been widely discussed in the literature whereas there has been few works on higher-order discretizations, with the exception of Refs. \cite{Coulaud2016,Carabias2018}.\\

Recently, the continuous mesh framework for metric-based mesh adaption \cite{Loseille2011I,Loseille2011II} has been extended to the higher-order case \cite{Coulaud2016,Carabias2018}. In the following, we briefly describe this framework and our proposed implementation. We denote by $s_i$ the component $i=1 \dots \dim{(s)}$ of the continuous sensor variable ($\dim{(s)} = 2$ in this work) and by $\Pi s_i$ its discrete P2 interpolation. The approximation error $||s_i-s_{h,i}||_{L^2(K)}$ is bounded within an element $K \in \mathcal{T}_h$ by the interpolation error $||s_i-\Pi s_i||_{L^2(K)}$ which is itself bounded by the $L^2$ norm of the third derivatives $||\nabla \nabla \nabla s_i||_{L^2(K)}$:
$$||s_i-\Pi s_i||_{L^2(K)} \leq \mathcal{C}(K) ||\nabla \nabla \nabla s_i||_{L^2(K)}, \forall K \in \mathcal{T}_h, \forall i = 1 \dots \dim{(s)}.$$
An approximation of the third derivatives of $s_i$ is recovered node-wise using SPR which is naturally extended for the P2 sensor variable by recovering the continuous first, second, third and fourth derivatives of $s_i$ around each node \cite{Belhamadia2004}. Hence, 15 unknowns are solved at each node in 2D and 35 in 3D. To ensure that this problem is well-defined, we use patches with at least twice as many nodes as unknowns for the least-squares fit. The node-wise continuous third derivatives of the two-dimensional sensor variable are extracted from this SPR operation.\\
The main challenge with respect to the P1 case is that $\nabla \nabla \nabla s_i$ is not a symmetric positive definite second-order tensor and thus does not define a quadratic form, a scalar product and a distance measure. The idea proposed in Ref. \cite{Coulaud2016} is to find a suitable symmetric positive definite matrix $\mathbf{Q}(\mathbf{x})$ at each point $\mathbf{x} \in \Omega$ such that
\begin{equation}\label{eq:adapt-bound}
|\nabla \nabla \nabla s_i(\mathbf{x}) . \mathbf{y} . \mathbf{y} . \mathbf{y}| \leq |\mathbf{y}^T \mathbf{Q}(\mathbf{x}) \mathbf{y}|^{\frac{3}{2}}, \forall \mathbf{y} \in \mathbb{R}^d, \forall i = 1 \dots \dim{(s)}.
\end{equation}
In order to have a close bound, the determinant of $\mathbf{Q}(\mathbf{x})$ should be as small as possible. This constrained minimization problem hence consists in finding a symmetric positive definite matrix $\mathbf{Q}(\mathbf{x})$ such that $\det{\mathbf{Q}(\mathbf{x})}$ is minimal and the bound in Eq. \eqref{eq:adapt-bound} is satisfied. It was solved in Ref. \cite{Coulaud2016} using a so-called logarithm transformation and a simplex minimization algorithm. Alternatively, a simpler approach consisting in directly finding a quadratic form approximating $\nabla \nabla \nabla s(\mathbf{x})$ through a least-squares fitting was proposed in Ref. \cite{Carabias2018}. In this work, we develop a similar approach which consists in directly approximating $\nabla \nabla \nabla s(\mathbf{x})$ as a quadratic form through the following geometric averaging operation:
\begin{equation}\label{eq:adapt-approx}
\mathbf{Q}(\mathbf{x}) = \left(\exp\left(\frac{1}{\dim{(s)}d} \sum_{i=1}^{\dim{(s)}} \sum_{j=1}^d \log\left(\left(\nabla \nabla \frac{\partial s_i}{\partial x_j}(\mathbf{x})\right)^{-\frac{1}{2}}\right)\right)\right)^{-2}.
\end{equation}
This geometric averaging operation is widely used to average or interpolate metric fields in various applications \cite{Arsigny2006,Laug2013}. It is called geometric because $\nabla \nabla \frac{\partial s_i}{\partial x_j}(\mathbf{x})$ is a symmetric positive definite matrix and can be associated to an ellipsoid whose radii are given by the eigenvalues of $(\nabla \nabla \frac{\partial s_i}{\partial x_j}(\mathbf{x}))^{-\frac{1}{2}}$. It is shown in Sec. \ref{sec:results} that this simple averaging operation preserves the directions embedded in $\nabla \nabla \nabla s(\mathbf{x})$ and, in particular, predicts correctly the decrease of the interpolation error in regions with flat interface.\\
The conversion of the directional errors carried by the metric field $\mathbf{Q}$ into a metric field $\mathbf{M}$ of directional mesh sizes requires the solution of a constrained minimization problem consisting in minimizing the total error while uniformly distributing local errors and controlling the complexity. The solution of this constrained minimization problem in the higher-order case can be expressed as \cite{Coulaud2016}
\begin{equation}\label{eq:adapt-metric}
\mathbf{M}(\mathbf{x}) = N^\frac{2}{d}\left(\int_\Omega (\det(\mathbf{Q}(\mathbf{x}))^\frac{3}{6+d} \mathrm{d}\mathbf{x}\right)^{-\frac{2}{d}} \left(\det(\mathbf{Q}(\mathbf{x}))\right)^{-\frac{1}{6+d}}\mathbf{Q}(\mathbf{x}),
\end{equation}
where $N$ is the prescribed number of P1 nodes. It is reminded that the asymptotic convergence rate in $L^2$ norm with respect to the mesh size for a uniform isotropic P2 mesh is 3, which means a rate of $\frac{3}{d}$ with respect to $N$.\\

The metric field defined in Eq. \eqref{eq:adapt-metric} can be used to compute the distorted lengths of each edge of the current FE mesh and determine whether they are too long (distorted length larger than one) or too short (distorted length smaller than one). The distorted volumes of each element of the current mesh can also be computed. An optimal mesh for a metric field $\mathbf{M}$ has a minimum number of elements whose distorted volumes are as uniform as possible and edges whose distorted lengths are as close as possible to one.\\
To solve this multi-objective optimization problem, we use a mesh adaption algorithm developed in previous works \cite{Gruau2005,Shakoor2017}. Note that we keep the P2 nodes at edge centers in this optimization and the unstructured mesh remains linear (only the interpolation is quadratic). For a given element, the quality criterion which is used in the algorithm considers the element's distorted volume and the average of the distorted lengths of the element's edges. All conformal topological modifications including node re-positioning and edge splitting are tested in small patches in the neighborhood of each node and edge of the mesh, and the modification which maximizes the quality criterion is accepted. This iterative algorithm with local modifications ends when no more local modification can be found to improve the quality of the mesh. The details on the mathematical background of this algorithm can be found in Ref. \cite{Gruau2005}, while the implementation details can be found in Ref. \cite{Shakoor2017}.\\
Once the mesh adaption algorithm terminates, the velocity field and the LS function should be transferred from the old mesh to the new (adapted) one. First, a space partitioning technique is used to locate efficiently the element of the old mesh containing each node of the new mesh. The values of the velocity and the LS function are then computed at each node of the new mesh using P2 interpolation from the containing element of the old mesh. At this step, the P2 discretization of both the velocity field and the LS function can reduce any errors and diffusion induced by mesh adaption and fields transfer. Finally, the new mesh and the transferred velocity field and LS function can be used for the current time increment.

\section{\label{sec:results}Results and discussion}

The proposed method has been implemented in a \textit{C} code (as of Ref. \cite{C99}), whereas the LS reinitialization procedure mentioned in Subsec. \ref{subsec:method:reinit} has been implemented in \textit{NVIDIA's CUDA} programming language for GPU computing and the mesh adaption algorithm presented in Subsec. \ref{subsec:method:adapt} has been implemented in \textit{C++} (as of Ref. \cite{Cpp11}). For computationally demanding operations such as numerical differentiation and integration of the discrete RBVMS weak form presented in Subsec. \ref{subsec:method:rbvms}, implementations rely on shared-memory multi-threaded  parallelism using \textit{OpenMP} \cite{OpenMP45}. The \textit{lis} library to solve the linear system at each iteration of the Newton-Raphson scheme also relies on \textit{OpenMP} \cite{Nishida2010}. In particular, we employ the generalized minimum residual method preconditioned using an incomplete LU factorization with threshold. The tolerance for this linear solver is set to $10^{-10}$, the perturbation for numerical differentiation $\delta_{ND}$ is set to $10^{-8}$, and the tolerance for the Newton-Raphson solver is set to $10^{-6}$.\\
Following the notations in Ref. \cite{Hysing2009}, we note NDOF the number of degrees of freedom for a given simulation. It is reminded that depending on the boundary conditions, there are at most $d+2$ DOF per node of the FE mesh. The number of time increments for a simulation is denoted as NTS, and the number of elements of the mesh as NEL.

\subsection{Single static inviscid droplet (3D)}

\begin{figure}[htbp]
	\centering
	\begin{subfigure}{0.49\textwidth}
	    \includegraphics[width=\textwidth,trim={165 95 360 140},clip]{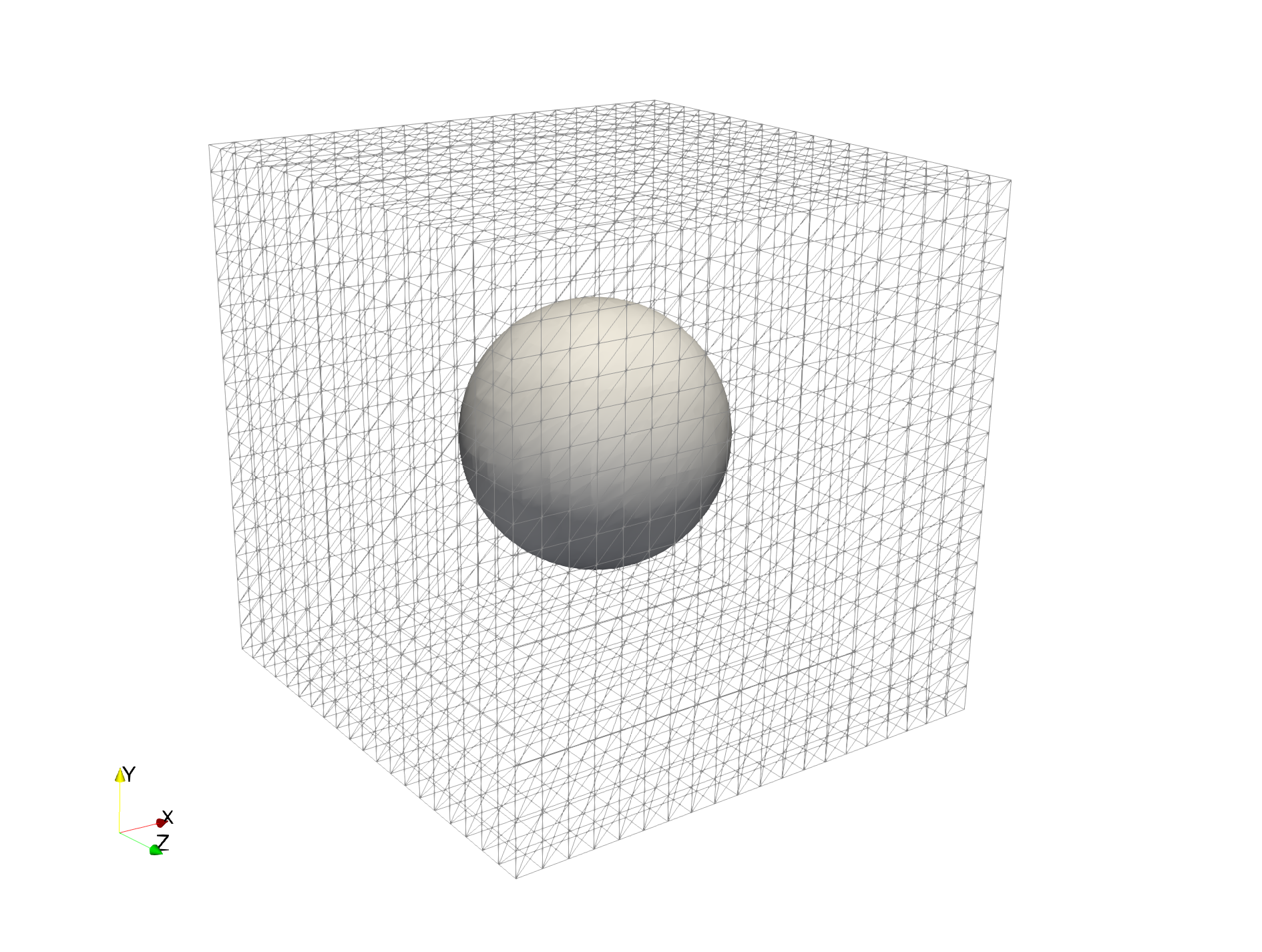}
		\caption{}
	\end{subfigure}%
	\begin{subfigure}{0.49\textwidth}
	    \includegraphics[width=\textwidth,trim={160 95 275 100},clip]{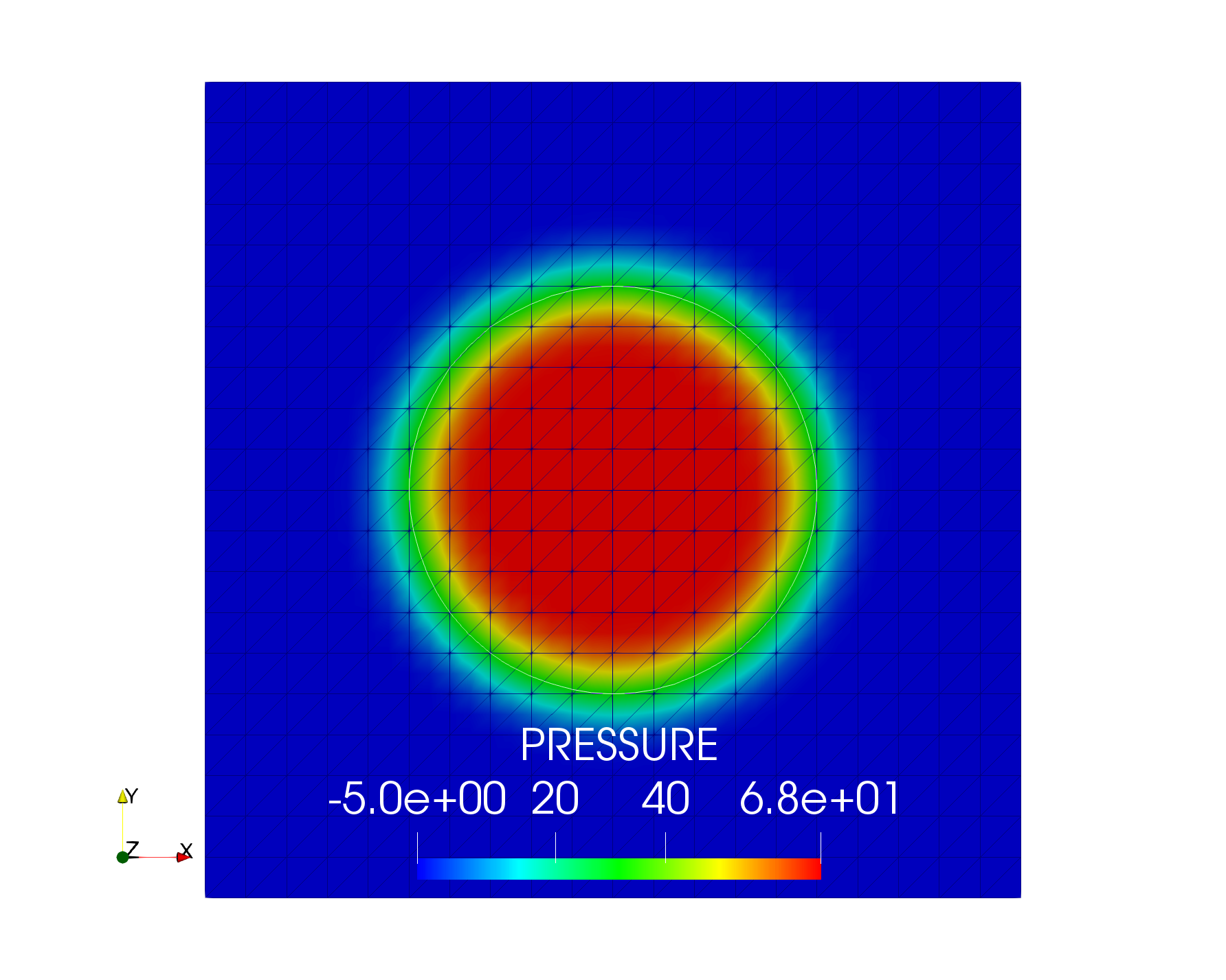}
		\caption{}
	\end{subfigure}%
\caption{3D single static inviscid droplet: (a) uniform $40^3$ mesh with droplet placed at its center, (b) pressure field computed at $t=0.050$ using the $20^3$ uniform P2 mesh and exact curvature.}
\label{fig:static_bubble_setup}
\end{figure}

This first set of simulations is conducted to assess the efficiency of the balanced-force implementation of the CSF model as well as the accuracy of curvature computation using P2 FEs in comparison with other works \cite{Williams1998,Francois2006,Yan2018iga,Lin2018}. Following these works, we define the computation domain as a cube of dimensions $8 \times 8 \times 8$, discretized with meshes of $40 \times 40 \times 40$ and $20 \times 20 \times 20$ elements. A spherical droplet of radius $R = 2$ is placed at the center of the computation domain as shown in Fig. \ref{fig:static_bubble_setup}(a) and no-slip conditions are applied at all the computation domain boundaries. Non-dimensionalized fluid and flow properties $\rho_1 = 1$, $\rho_2 = 10^{-1}$, $\mu_1 = \mu_2 = 0$, $g = 0$, $\sigma_s = 73$ are used. The simulation is conducted for $50$ time increments using a time step $\Delta t = 0.001$. Under these conditions, the inviscid droplet should remain static and the pressure difference $\Delta p$ between outside and inside the droplet should converge to that given by the Young-Laplace equation:
$$\Delta p = \sigma_s \kappa_s = 73,\text{ with }\kappa_s = \frac{2}{R} = 1.$$
As an error measure, we use the maximum Euclidean norm of the velocity vector in the domain $\max_{\mathbf{x} \in \Omega}(||v(\mathbf{x},t)||_2)$, after one time increment and after 50. At the same instants, we also report the error of the difference between the highest and the lowest pressures Error($\Delta p_{max}(t)$).\\

First, using the exact curvature in the numerical scheme, both errors should be below the tolerance used for the Newton-Raphson scheme, owing to the balanced-force implementation of the CSF model. In this case, we can also conduct the simulation using equal-order P1 FEs, because we do not need to compute the curvature. After 50 time increments, errors might increase depending on the accuracy of the numerical method, in particular the LS reinitialization method.\\
Results are reported in the first part of Tab. \ref{tab:static-droplet}. At equal NDOF, both errors of pressure and of velocity are below the tolerance used for the Newton-Raphson scheme, even after 50 time increments, regardless of the order of the FE method.  This result is in agreement with the theory and validates our balanced-force implementation. Contrary to other works \cite{Lin2018}, we cannot reduce this error down to machine precision, which is mainly due to the numerical differentiation that uses $\delta_{ND} = 10^{-8}$.\\
Note that in Tab. \ref{tab:static-droplet} we have also reported the regularization thickness $\epsilon$. There can be significant oscillations of the pressure field near the interface if $\epsilon$ is much smaller than the mesh size. For instance, the pressure field obtained using exact curvature and a value of $\epsilon$ corresponding to a regularization over four elements is shown in Fig. \ref{fig:static_bubble_setup}(b).\\

\begin{table}[htbp]
    \centering
    \begin{tabular}{r||c|c||c|c}
        Mesh & $40^3$ & $20^3$ & $40^3$ & $20^3$ \\
        $\epsilon$ & $0.4$ & $0.8$ & $0.4$ & $0.8$ \\
        Curvature & Exact & Exact & Numerical & Numerical \\
        FE & P1 & P2 & P2 & P2 \\
        NDOF & 315,799 & 315,799 & 2,541,999 & 315,799 \\ \hline
        Error($\Delta p_{max}(t=0.001)$) & $< 10^{-8}$ & $< 10^{-8}$ & $5.27 \times 10^{-1}$ & $1.44 \times 10^{0}$ \\
        Error($\Delta p_{max}(t=0.050)$) & $1.64 \times 10^{-7}$ & $3.60 \times 10^{-8}$ & $1.31 \times 10^{1}$ & $2.72 \times 10^{0}$ \\
        $\max_{\mathbf{x} \in \Omega}(||v(\mathbf{x},t=0.001)||_2)$ & $< 10^{-8}$ & $< 10^{-8}$ & $3.33 \times 10^{-2}$ & $2.32 \times 10^{-2}$ \\
        $\max_{\mathbf{x} \in \Omega}(||v(\mathbf{x},t=0.050)||_2)$ & $8.61 \times 10^{-8}$ & $4.43 \times 10^{-8}$ & $1.07 \times 10^{1}$ & $1.75 \times 10^{0}$
    \end{tabular}
    \caption{3D single static inviscid droplet simulation results for different FE meshes and orders, using either exact or numerical curvature. Errors below $\delta_{ND} = 10^{-8}$ are not reported.}
    \label{tab:static-droplet}
\end{table}

Second, using numerically computed curvature, errors will depend on the interpolation error of the LS function and its second derivatives. In the literature, most authors rely on derivatives recovery techniques which can smooth the mean curvature \cite{Williams1998,Francois2006,Zahedi2012,Lin2018,Yan2018am}. A higher-order and higher-regularity isogeometric analysis method has also been proposed \cite{Yan2018iga}. Here, we use a P2 discretization of the LS function without derivatives recovery nor smoothing technique.\\
Results are reported in the second part of Tab. \ref{tab:static-droplet}. There are clearly numerical errors in the computation of the mean curvature, which may significantly affect simulation results depending on the discretization at the vicinity of the interface. These errors accumulate with time steps and are hence higher after 50 steps. This error accumulation effect seems to be alleviated with a second order time stepping scheme \cite{Yan2018iga} which could be worth considering.\\
In the literature \cite{Williams1998}, various derivatives recovery and smoothing techniques using the finite difference method have been reported to lead to errors of the velocity after one time increment between $8.55 \times 10^{-2}$ and $3.49 \times 10^{-1}$ for the $40^3$ mesh. After 50 time increments, those errors have been reported to vary between $3.86 \times 10^{-1}$ and $2.55 \times 10^{0}$ depending on the smoothing technique. These works did not employ a balanced-force algorithm. For example, the velocity errors between $4.02 \times 10^{-3}$ and $4.87 \times 10^{-3}$ after one time increment, and between $4.02 \times 10^{-2}$ and $1.63 \times 10^{-1}$ after 50 time increments have been reported in Ref. \cite{Francois2006} using the finite difference method with different smoothing techniques for the $40^3$ mesh. Overall, our results after one time increment are better than those obtained with non balanced-force implementations, whereas they are not as accurate as those reported in Ref. \cite{Francois2006}. It seems that the balanced-force implementation is necessary even though it is not sufficient to eliminate spurious currents. Thus, the balanced-force implementation should be combined with curvature smoothing techniques.\\
This speculation is confirmed by the fact that our errors are at least two orders superior to those reported in Ref. \cite{Yan2018iga} using equal-order P1 FEs and a projection technique to recover the mean curvature. Although such technique would make the numerical differentiation in our fully implicit scheme non-local and hence more difficult to implement, it may be worth considering in the future.\\
Ref. \cite{Yan2018iga} reported that isogeometric analysis with its higher-order basis functions led to a more accurate mean curvature computation but did not lead to a reduction of spurious currents, compared to a P1 discretization. This is also confirmed by the results using numerical curvature in Tab. \ref{tab:static-droplet}.\\
However, higher-order methods are still relevant to improve the accuracy on the pressure field. Errors after 50 time increments of $1.39 \times 10^{0}$ for the $40^3$ mesh and $3.93 \times 10^{0}$ for the $20^3$ mesh have been reported in Ref. \cite{Yan2018iga} using the equal-order P1 FE method with projection, and $7.90 \times 10^{-1}$ and $1.37 \times 10^{0}$ using isogeometric analysis. Our results are of the same order but we observe a drastic increase of the error after 50 time increments, which emphasizes again the need to improve the time stepping scheme in future developments.\\

As a conclusion, this first set of simulations validates our balanced-force implementation of the CSF model, because all computations using exact mean curvature lead to errors both in velocity and pressure which are below the tolerance used for the Newton-Raphson solver. Regarding computations with numerically computed mean curvature, spurious non-physical currents are not completely eliminated. However, the proposed equal-order P2 framework enables a good compromise between the accuracy of the pressure field and the simplicity of the scheme because we do not use any projection or curvature smoothing technique.

\subsection{Single bubble rise (2D)}

\begin{figure}[htbp]
	\centering
	\includegraphics[width=0.4\textwidth]{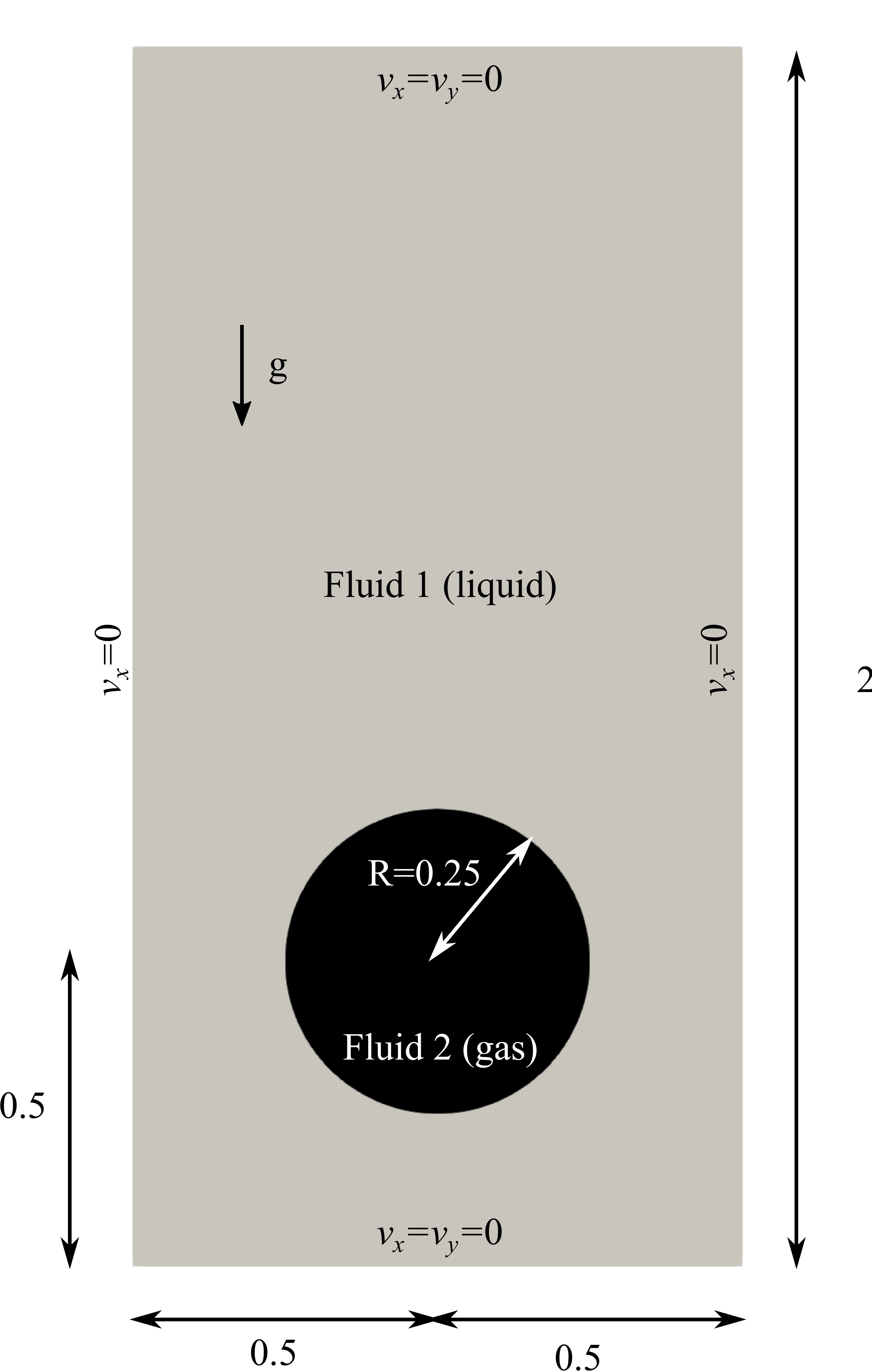}
\caption{Setup for the 2D single bubble rise simulations.}
\label{fig:bubble_rise_2D_setup}
\end{figure}

This second set of simulations considers the dynamics of bubble rise in a 2D domain. The simulation setup shown in Fig. \ref{fig:bubble_rise_2D_setup} has been proposed in a benchmark for multiphase flow simulation methods \cite{Hysing2009}. The computation domain is a rectangle of dimensions $1 \times 2$, discretized with meshes of $40 \times 80$,  $80 \times 160$ and  $160 \times 320$. A circular bubble of radius $R = \REV{0.25}$ is initially placed at $(0.5,0.5)$. Free-slip boundary conditions are applied at lateral boundaries, while no-slip conditions are applied at top and bottom boundaries. Two cases corresponding to two sets of non-dimensionalized fluid and flow properties \cite{Hysing2009} are used:
\begin{itemize}
\item for case 1, properties are $\rho_1 = 1000$, $\rho_2 = 100$, $\mu_1 = 10$, $\mu_2 = 1$, $g = 0.98$, $\sigma_s = 24.5$,
\item and for case 2, $\rho_1 = 1000$, $\rho_2 = 1$, $\mu_1 = 10$, $\mu_2 = 0.1$, $g = 0.98$, $\sigma_s = 1.96$.
\end{itemize}
\REV{Case 1 corresponds to the Reynolds number of Re=35 and the E\"otv\"os number of Eo=10, while case 2 corresponds to Re=35 and Eo=125. For the definitions of these dimensionless numbers in this particular setup, the reader is referred to Ref. \cite{Hysing2009}.}
For each case, we measure and report the relative $L^2$ errors of the bubble area $V_b$, the bubble center of mass $y$ coordinate $y_{b}$ and the bubble rise velocity $v_{b}$ integrated over time. Those parameters are respectively given by
$$V_{b}(t) = \int_\Omega H(-\phi(\mathbf{x},t)) \mathrm{d}\Omega,$$
$$y_{b}(t) = \frac{\int_\Omega H(-\phi(\mathbf{x},t)) \mathbf{x} \mathrm{d}\Omega}{V_{b}(t)}.\mathbf{e}_y,$$
$$v_{b}(t) = \frac{\int_\Omega H(-\phi(\mathbf{x},t)) \mathbf{v}(t) \mathrm{d}\Omega}{V_{b}(t)}.\mathbf{e}_y.$$
The relative $L^2$ error integrated over time is
$$\text{Error}(q) = \sqrt{\frac{\int_0^T (q(t) - q_{ref}(t))^2 \mathrm{d}t}{\int_0^T (q_{ref}(t))^2 \mathrm{d}t}},$$
where $T=3$ and $q=V_b, y_b,\text{ or }v_b$. Simulations with different space (NDOF) and time (NTS) discretizations are conducted with and without anisotropic mesh adaption. The reference solution $V_{b,ref}$ is $\pi R^2$, as the flow is incompressible. Since we do not have analytical solutions for $y_{b,ref}$ and $v_{b,ref}$, we use the numerical solutions corresponding to the simulation with best $V_{b}$ accuracy for each case.\\
The asymptotic convergence rate with respect to the square root of NDOF is 3, as explained in Subsec. \ref{subsec:method:adapt}, while the asymptotic convergence rate with respect to NTS is 1. Even though the numerical scheme is unconditionally stable, these optimal convergence rates will not be reached if the mesh size is too small compared to the time step. In Ref. \cite{Hysing2009}, convergence rates corresponded to a simultaneous reduction of the mesh size and the time step. Furthermore, for a given NDOF, increasing NTS may not necessarily increase the accuracy as more time increments mean more LS reinitializations. Indeed, the LS reinitialization method (Subsec. \ref{subsec:method:reinit}) slightly shifts the interface at each reinitialization.

\subsubsection{Case 1}

\begin{figure}[htbp]
	\centering
	\begin{subfigure}{0.3\textwidth}
	    \includegraphics[height=9cm,trim={320 30 900 80},clip]{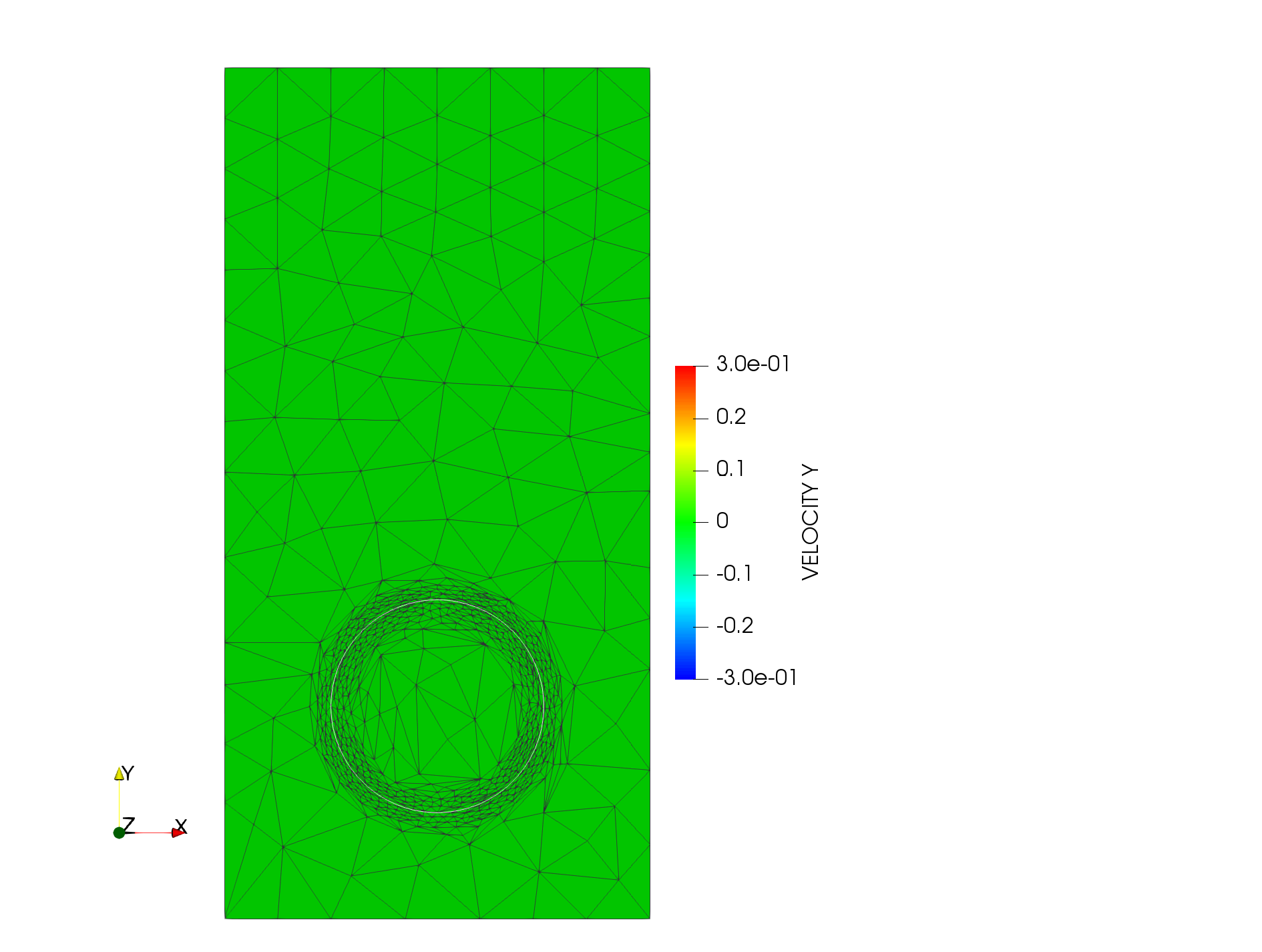}
		\caption{}
	\end{subfigure}%
	\begin{subfigure}{0.3\textwidth}
	    \includegraphics[height=9cm,trim={320 30 900 80},clip]{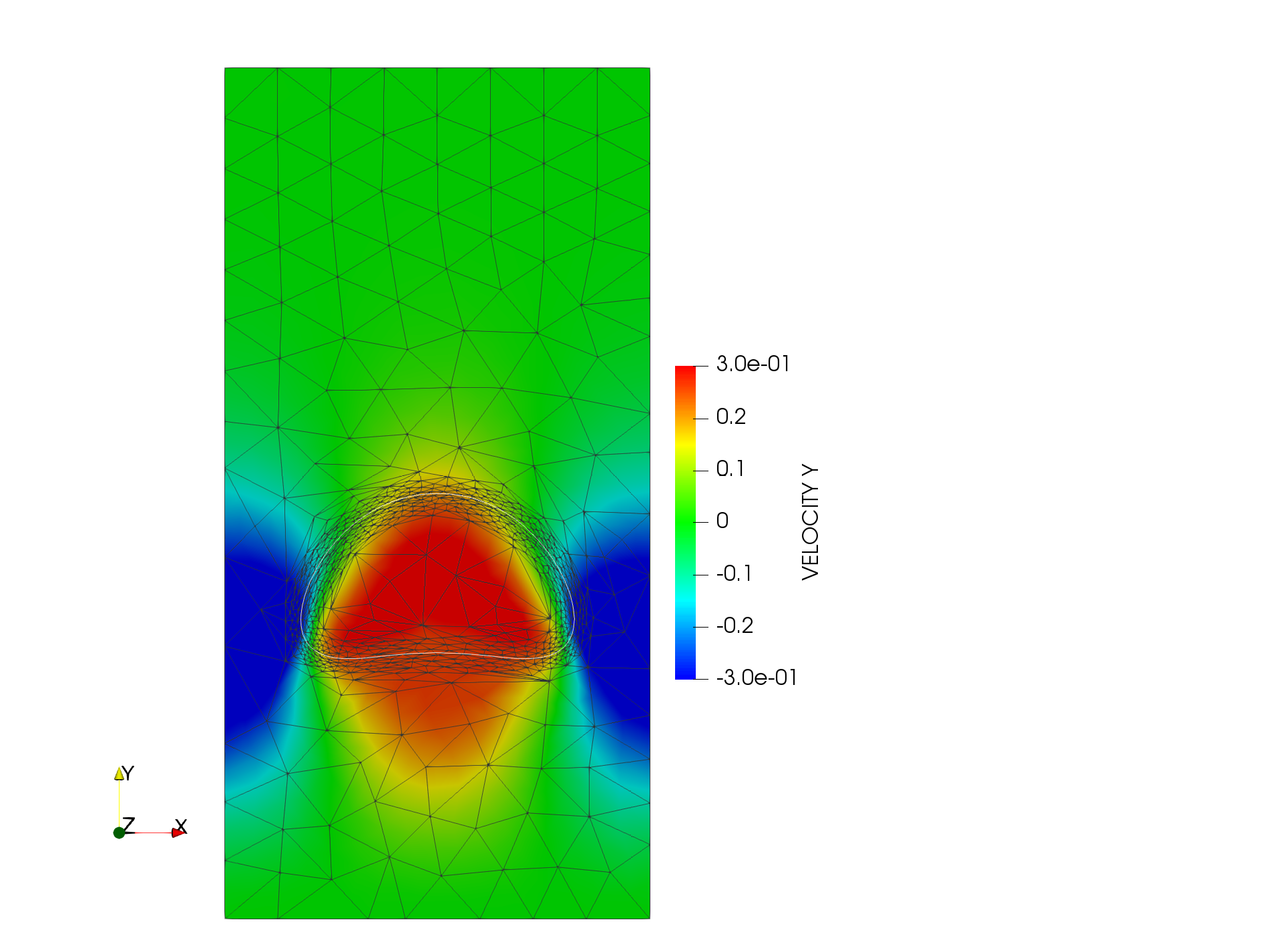}
		\caption{}
	\end{subfigure}%
	\begin{subfigure}{0.4\textwidth}
	    \includegraphics[height=9cm,trim={320 30 650 80},clip]{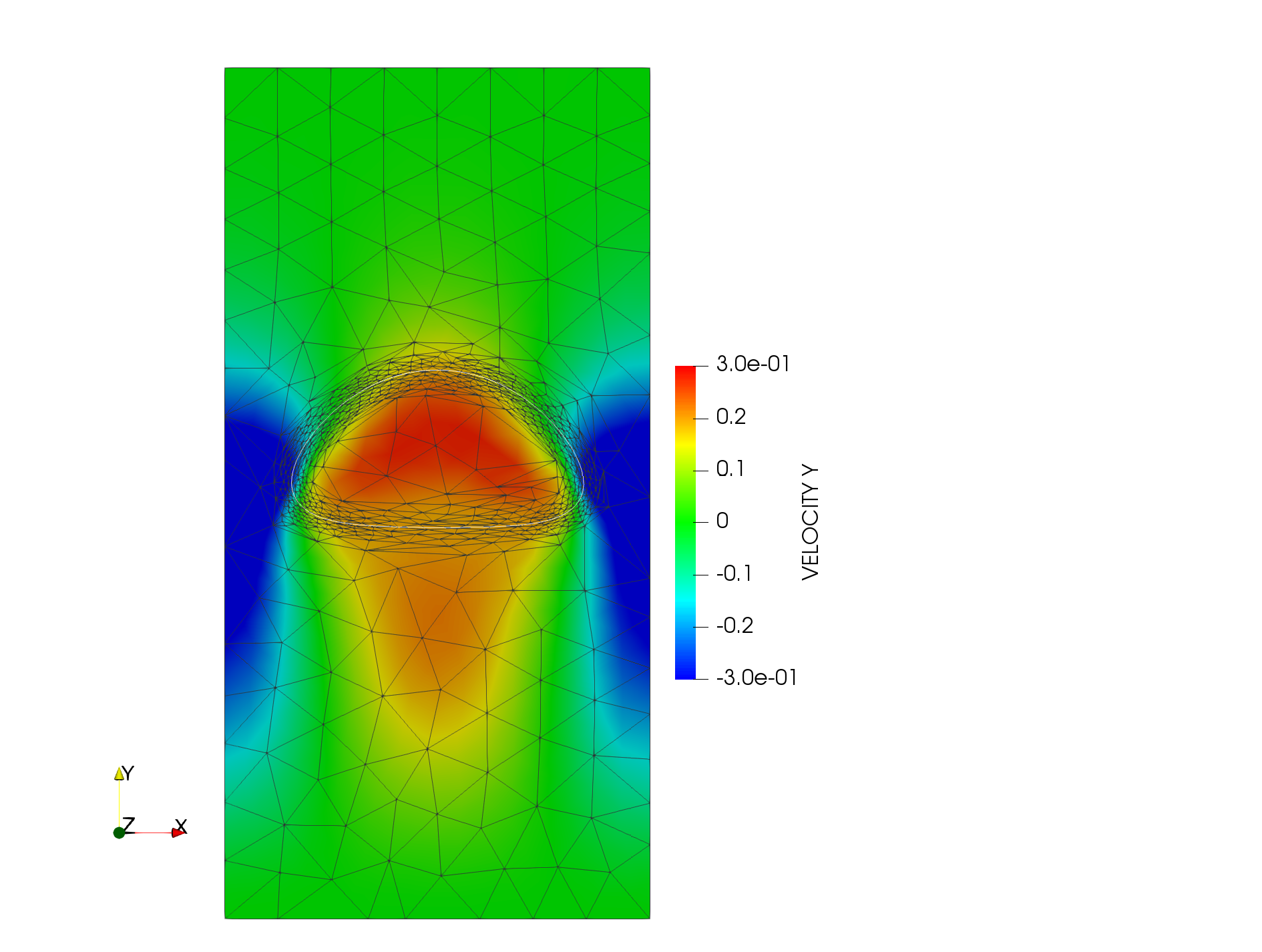}
		\caption{}
	\end{subfigure}%
\caption{Velocity field in rise direction $\mathbf{v}.\mathbf{e}_y$ for the 2D single bubble rise case 1 simulation using an adaptive mesh with NDOF $\approx 11500$ and NTS $=407$ at: (a) $t = 0$, (b) $t = 1.5$, (c) $t=3.0$.}
\label{fig:bubble_rise_2D_case_1_result}
\end{figure}

\begin{figure}[htbp]
	\centering
	\includegraphics[width=0.8\textwidth]{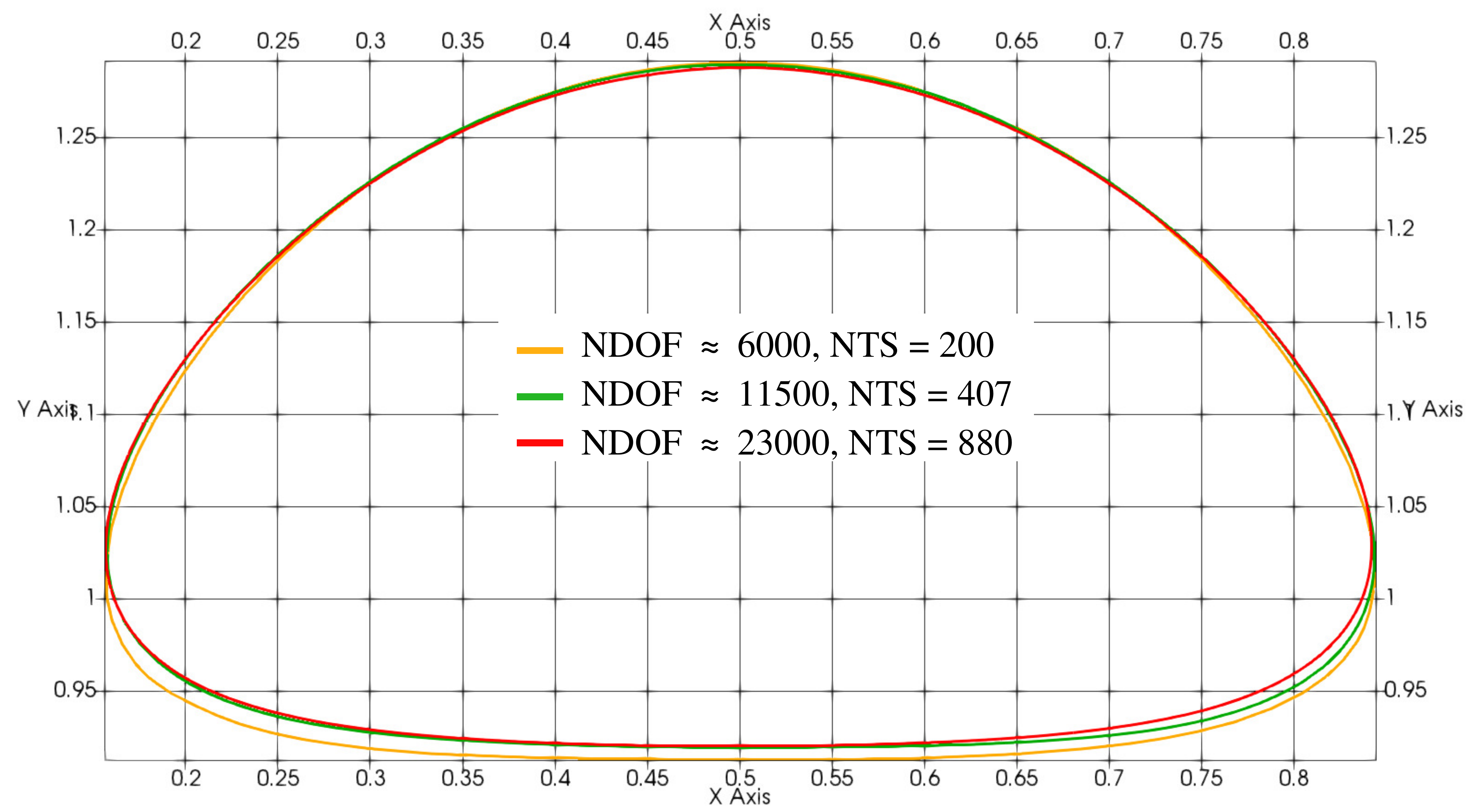}
\caption{Liquid-gas interface for some of the 2D single bubble rise case 1 simulations with adaptive mesh at $t=3$.}
\label{fig:bubble_rise_2D_case_1_shape}
\end{figure}

In case 1, the bubble which was initially circular becomes nearly elliptical with a flat bottom surface. One of our solutions computed using an adaptive mesh is shown in Fig. \ref{fig:bubble_rise_2D_case_1_result}. Anisotropic mesh adaption is particularly relevant for this kind of interface morphology because elements can be stretched along directions where the third derivatives of the LS function are close to zero, to reduce the computational cost. This can be seen in Fig. \ref{fig:bubble_rise_2D_case_1_result}(c) in the lower part of the bubble. This anisotropic stretching of elements is also relevant to control the computational cost, because there is a longer perimeter to cover in Fig. \ref{fig:bubble_rise_2D_case_1_result}(c) than in Fig. \ref{fig:bubble_rise_2D_case_1_result}(a), while the number of elements is almost the same, as explained in Subsec. \ref{subsec:method:adapt}.\\
A closer look at the final shapes of the bubble for some of the simulations with adaptive mesh is presented in Fig. \ref{fig:bubble_rise_2D_case_1_shape}. A good convergence is obtained by dint of the fine mesh which is maintained near the interface, using anisotropic mesh adaption.\\

Errors in the computation of the parameters of interest are reported in Tab. \ref{tab:bubble-rise-2D-case-1-errors}. Although NTS was prescribed to be equal to 100, 200, 400 and 800, the Newton-Raphson problem was too stiff for the simulations with very fine meshes for the given time step. Thus, the time step was automatically decreased to obtain a solution as explained in Subsec. \ref{subsec:method:rbvms}. The most extreme case was the simulation with uniform $80 \times 160$ mesh and a prescribed NTS of 100 which was automatically increased to nearly 200. Results from this simulation were not reported in Tab. \ref{tab:bubble-rise-2D-case-1-errors} to avoid any confusion.

\begin{table}[htbp]
    \centering
    \begin{tabular}{r||c|c|c||c|c|c}
        Mesh & $20 \times 40$ & $40 \times 80$ & $80 \times 160$ & Adaptive & Adaptive & Adaptive \\
        $\epsilon$ & $0.1$ & $0.05$ & $0.025$ & $0.05$ & $0.025$ & $0.0125$ \\
        NEL & 1,600 & 6,400 & 25,600 & $\approx$ 740 & $\approx$ 1,400 & $\approx$ 2,800 \\
        NDOF & 12,962 & 51,522 & 205,442 & $\approx$ 6,000 & $\approx$ 11,500 & $\approx$ 23,000 \\ \hline
        NTS & 102 & 100 &  & 100 & 104 & 165 \\
        Error($V_b$) & $5.45 \times 10^{-2}$ & $4.79 \times 10^{-2}$ &  & $4.89 \times 10^{-2}$ & $4.68 \times 10^{-2}$ & $4.32 \times 10^{-2}$ \\
        Error($y_b$) & $4.45 \times 10^{-3}$ & $2.46 \times 10^{-3}$ &  & $2.13 \times 10^{-3}$ & $1.81 \times 10^{-3}$ & $3.41 \times 10^{-3}$ \\
        Error($v_b$) & $8.05 \times 10^{-2}$ & $3.01 \times 10^{-2}$ &  & $2.74 \times 10^{-2}$ & $1.74 \times 10^{-2}$ & $1.62 \times 10^{-2}$ \\ \hline
        NTS & 207 & 200 & 200 & 200 & 204 & 261 \\
        Error($V_b$) & $3.30 \times 10^{-2}$ & $2.40 \times 10^{-2}$ & $2.33 \times 10^{-2}$ & $2.05 \times 10^{-2}$ & $2.13 \times 10^{-2}$ & $2.14 \times 10^{-2}$ \\
        Error($y_b$) & $4.89 \times 10^{-3}$ & $2.51 \times 10^{-3}$ & $1.95 \times 10^{-3}$ & $2.99 \times 10^{-3}$ & $\mathbf{7.48 \times 10^{-4}}$ & $1.32 \times 10^{-3}$ \\
        Error($v_b$) & $7.64 \times 10^{-2}$ & $2.44 \times 10^{-2}$ & $1.37 \times 10^{-2}$ & $2.38 \times 10^{-2}$ & $\mathbf{9.25 \times 10^{-3}}$ & $9.59 \times 10^{-3}$ \\ \hline
        NTS & 406 & 400 & 400 & 400 & 407 & 463 \\
        Error($V_b$) & $2.71 \times 10^{-2}$ & $1.06 \times 10^{-2}$ & $1.10 \times 10^{-2}$ & $2.25 \times 10^{-3}$ & $6.18 \times 10^{-3}$ & $8.22 \times 10^{-3}$ \\
        Error($y_b$) & $7.45 \times 10^{-3}$ & $2.24 \times 10^{-3}$ & $1.85 \times 10^{-3}$ & $1.52 \times 10^{-3}$ & $\mathbf{4.51 \times 10^{-4}}$ & $1.86 \times 10^{-3}$ \\
        Error($v_b$) & $8.39 \times 10^{-2}$ & $2.12 \times 10^{-2}$ & $1.05 \times 10^{-2}$ & $1.60 \times 10^{-2}$ & $\mathbf{5.07 \times 10^{-3}}$ & $8.04 \times 10^{-3}$ \\ \hline
        NTS & 802 & 800 & 800 & 800 & 804 & 880 \\
        Error($V_b$) & $3.10 \times 10^{-2}$ & $\mathbf{1.99 \times 10^{-3}}$ & $3.92 \times 10^{-3}$ & $2.11 \times 10^{-2}$ & $\mathbf{7.17 \times 10^{-3}}$ & $\mathbf{1.67 \times 10^{-3}}$ \\
        Error($y_b$) & $1.03 \times 10^{-2}$ & $\mathbf{9.72 \times 10^{-4}}$ & $2.05 \times 10^{-3}$ & $1.49 \times 10^{-3}$ & $9.10 \times 10^{-4}$ & \\
        Error($v_b$) & $7.80 \times 10^{-2}$ & $1.71 \times 10^{-2}$ & $9.20 \times 10^{-3}$ & $1.48 \times 10^{-2}$ & $7.09 \times 10^{-3}$ &
    \end{tabular}
    \caption{Relative $L^2$ errors for 2D single bubble rise case 1 simulations using different space (NDOF) and time (NTS) discretizations, with and without anisotropic mesh adaption. All results with convergence rate with respect to the square root of NDOF higher than 2.5 are highlighted in bold.}
    \label{tab:bubble-rise-2D-case-1-errors}
\end{table}

As shown in the left part of Tab. \ref{tab:bubble-rise-2D-case-1-errors} where no mesh adaption was used, there is a compromise to find between the time discretization and the space discretization. Optimal convergence rates in $V_b$ and $y_b$ are reached for instance for the intermediate $40 \times 80$ mesh using the finest time discretization (highlighted in bold in Tab. \ref{tab:bubble-rise-2D-case-1-errors}). This confirms that if the mesh is not fine enough, interface shifts due to LS reinitialization become the dominant source of error and lead to an increased error with smaller time steps.\\
Overall good mass conservation is obtained because of the use of a P2 LS function. This is drastically improved using an adaptive mesh. For instance, using an adaptive mesh with NDOF $\approx 6,000$ (right part of Tab. \ref{tab:bubble-rise-2D-case-1-errors}), the error of $V_b$ for NTS = 400 is lower than the errors obtained using a uniform mesh with same NTS, even using the finest uniform mesh. A factor of reduction in terms of NDOF close to 35 is hence achieved.\\
Anisotropic mesh adaption is also relevant to reach optimal convergence rates for a lower NDOF than uniform meshes. This can be seen in the last line (NTS $\approx$ 800) of Tab. \ref{tab:bubble-rise-2D-case-1-errors} where an optimal convergence rate on $V_b$ is obtained using an adaptive mesh with NDOF $\approx 11,500$. Using an adaptive mesh with NDOF $\approx 11,500$ and NTS = 204, all errors are lower than the errors obtained using a uniform mesh with same NTS, which results in a factor of reduction in terms of NDOF close to 15.\\
For case 1, there was very good agreement between the simulation results obtained with the different computational methods assessed in the benchmark \cite{Hysing2009}. This is also verified for our results. A more thorough comparison is proposed in the following for case 2 where significantly different results were obtained in the benchmark depending on the computational method.

\subsubsection{Case 2}

\begin{figure}[htbp]
	\centering
	\begin{subfigure}{0.3\textwidth}
	    \includegraphics[height=9cm,trim={320 30 900 80},clip]{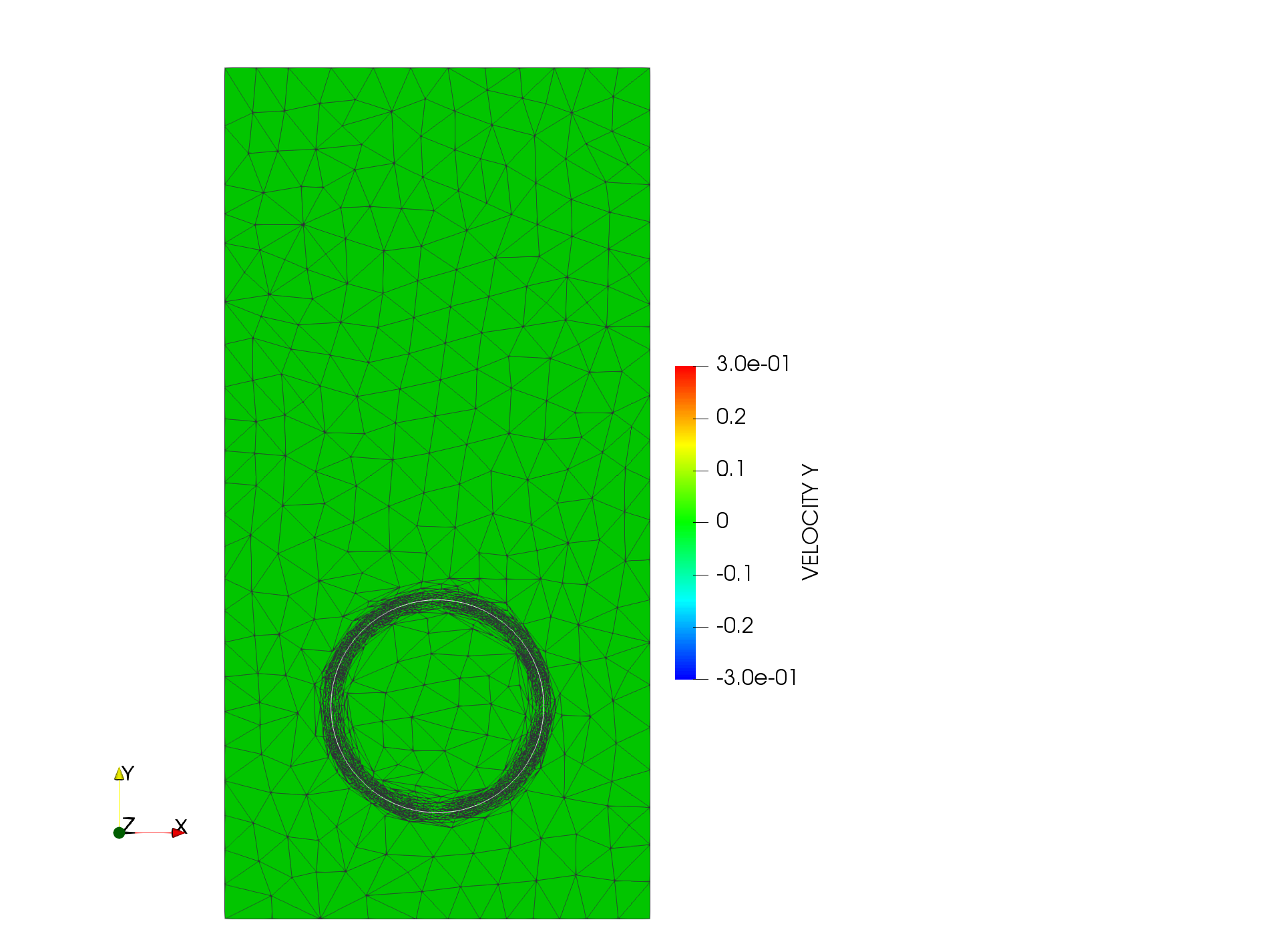}
		\caption{}
	\end{subfigure}%
	\begin{subfigure}{0.3\textwidth}
	    \includegraphics[height=9cm,trim={320 30 900 80},clip]{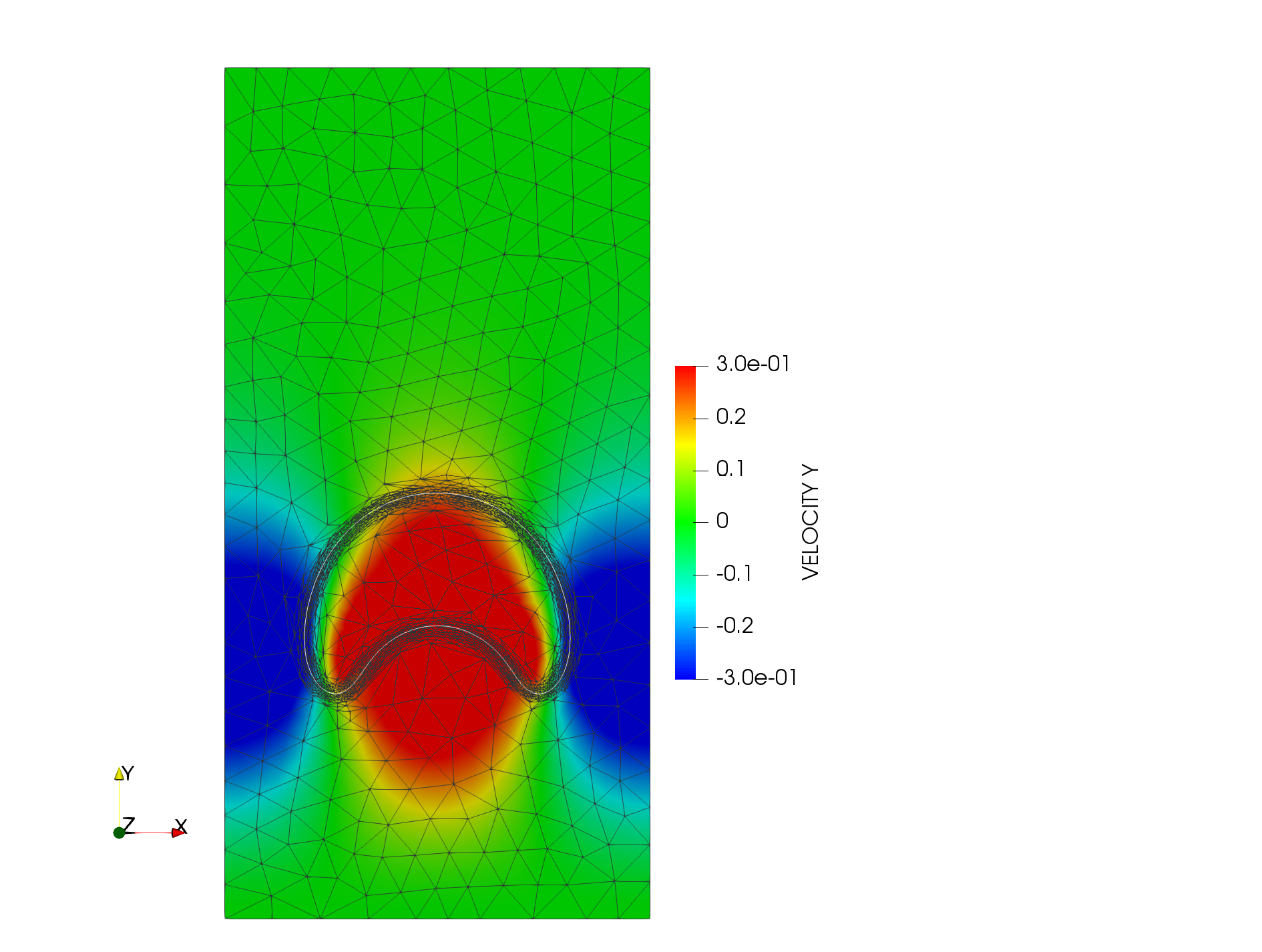}
		\caption{}
	\end{subfigure}%
	\begin{subfigure}{0.4\textwidth}
	    \includegraphics[height=9cm,trim={320 30 300 80},clip]{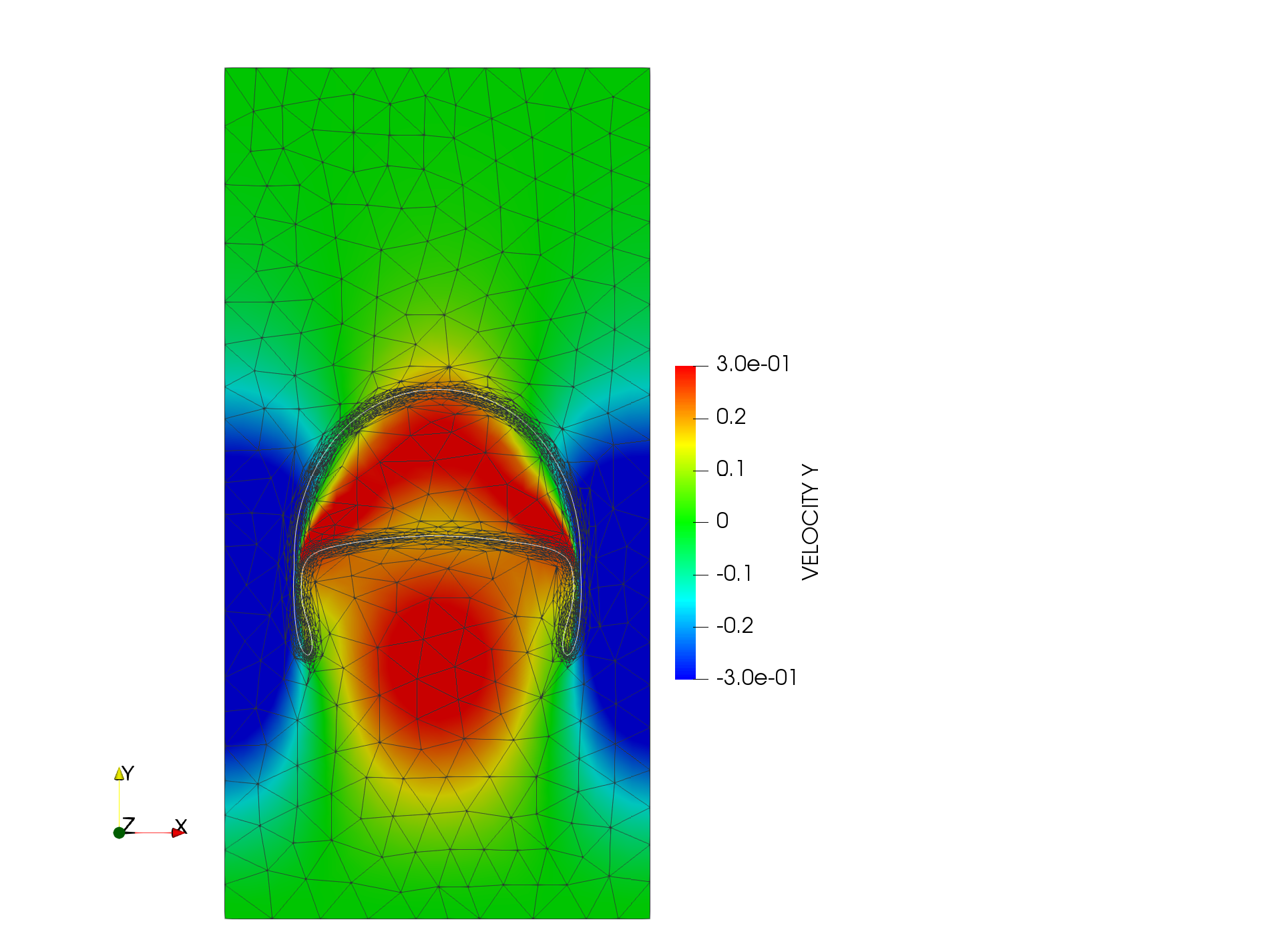}
		\caption{}
	\end{subfigure}%
	\hfill
	\begin{subfigure}{0.3\textwidth}
	    \includegraphics[width=\textwidth,trim={50 50 50 50},clip]{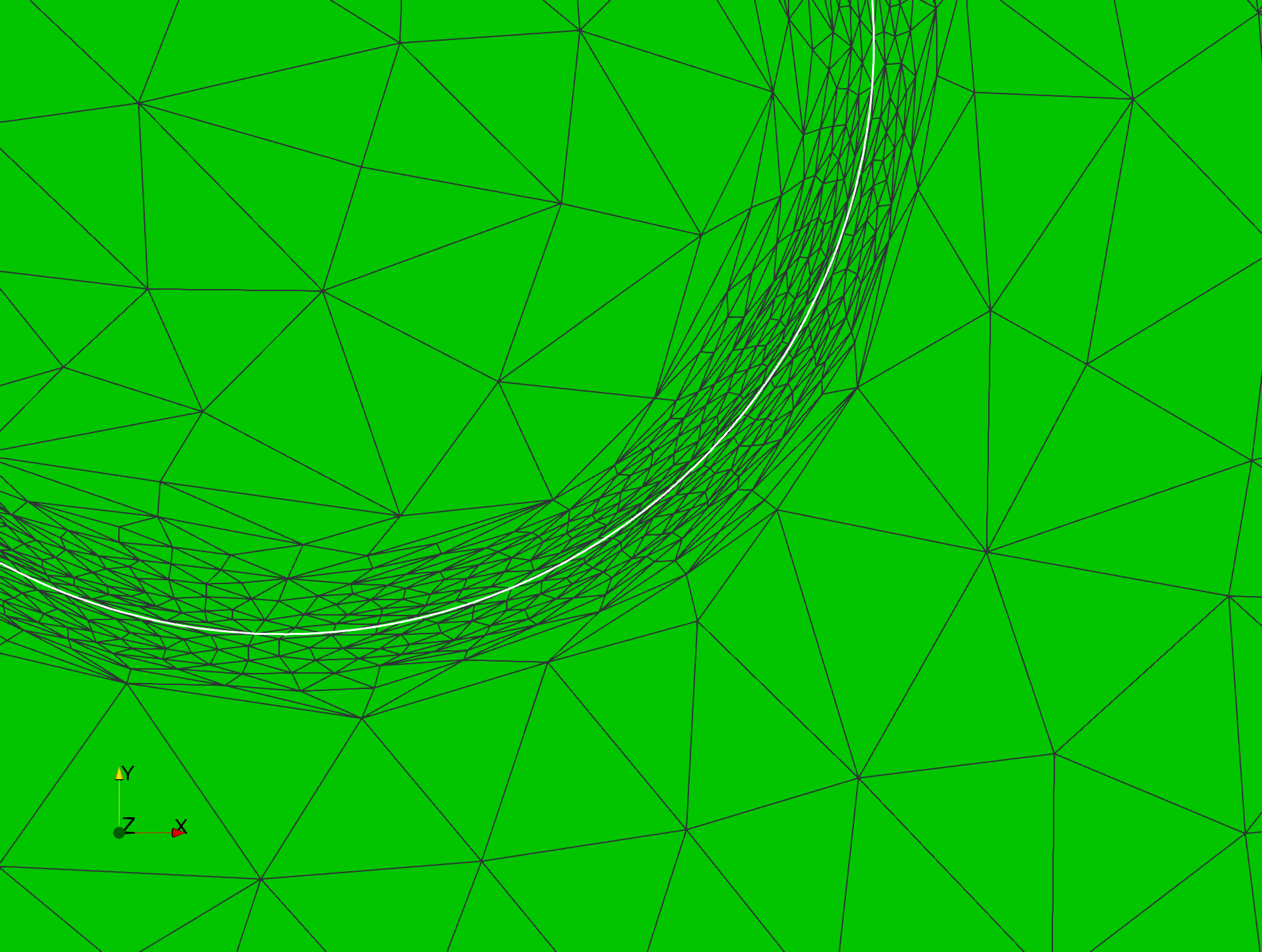}
		\caption{}
	\end{subfigure}%
	~
	\begin{subfigure}{0.3\textwidth}
	    \includegraphics[width=\textwidth,trim={50 50 50 50},clip]{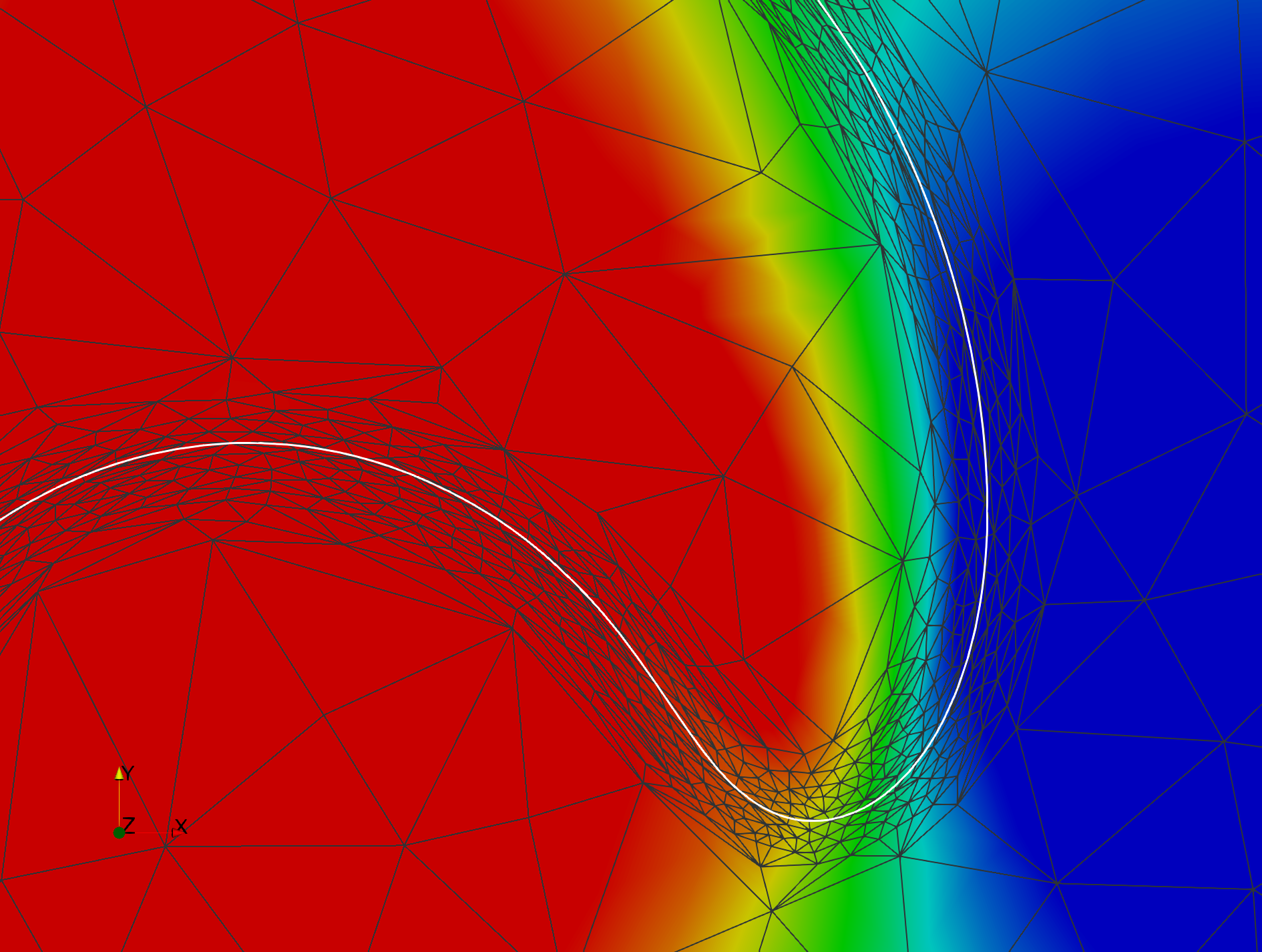}
		\caption{}
	\end{subfigure}%
	~
	\begin{subfigure}{0.3\textwidth}
	    \includegraphics[width=\textwidth,trim={50 50 50 50},clip]{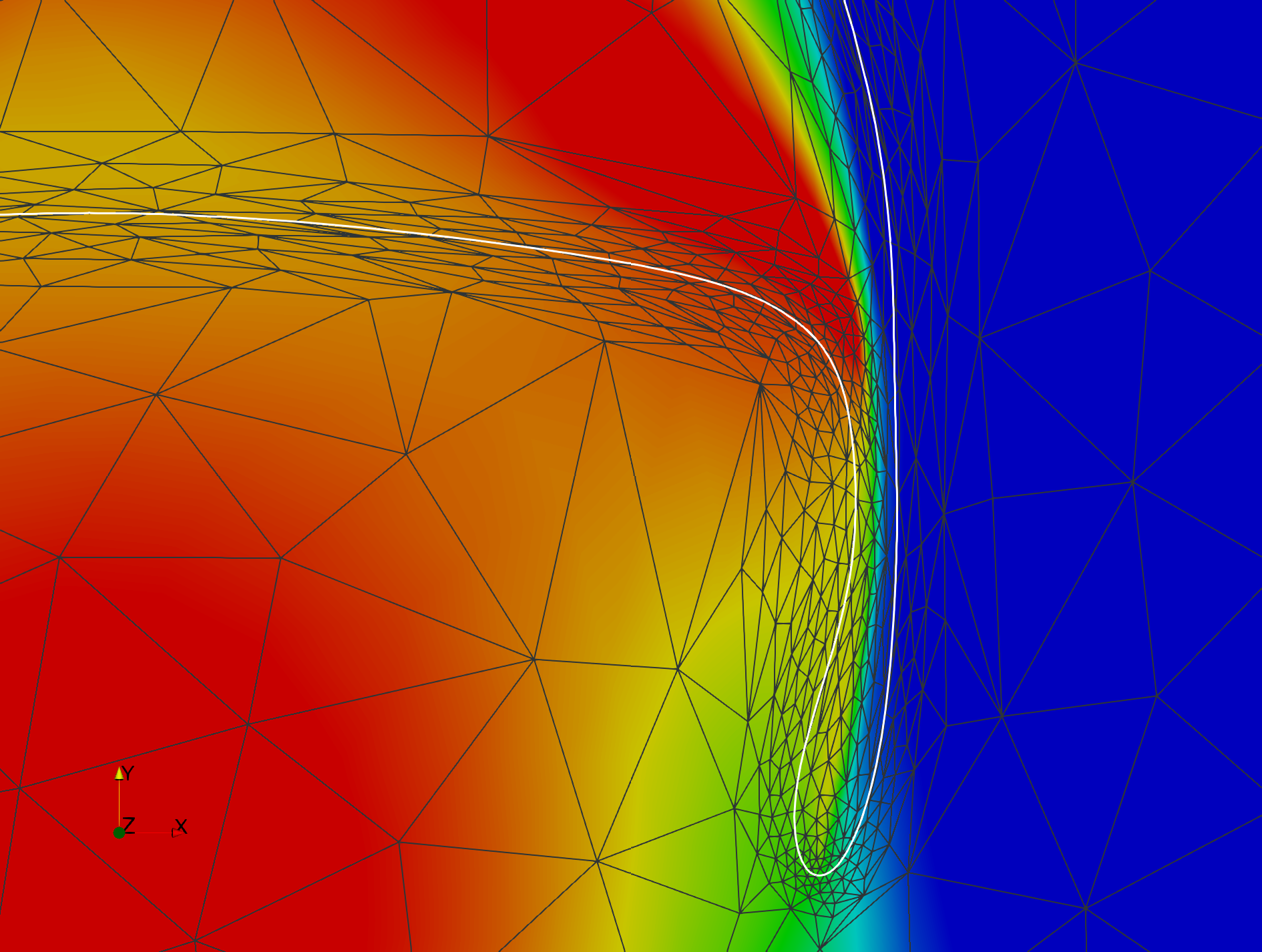}
		\caption{}
	\end{subfigure}%
\caption{Velocity field in rise direction $\mathbf{v}.\mathbf{e}_y$ for the 2D single bubble rise case 2 simulation using an adaptive mesh with NDOF $\approx 23000$ and NTS $=407$ at: (a) $t = 0$, (b) $t = 1.5$, (c) $t=3.0$. Zoomed images on the adapted mesh at: (d) $t = 0$, (e) $t = 1.5$, (f) $t=3.0$.}
\label{fig:bubble_rise_2D_case_2_result}
\end{figure}

\begin{figure}[htbp]
	\centering
	\includegraphics[width=0.8\textwidth]{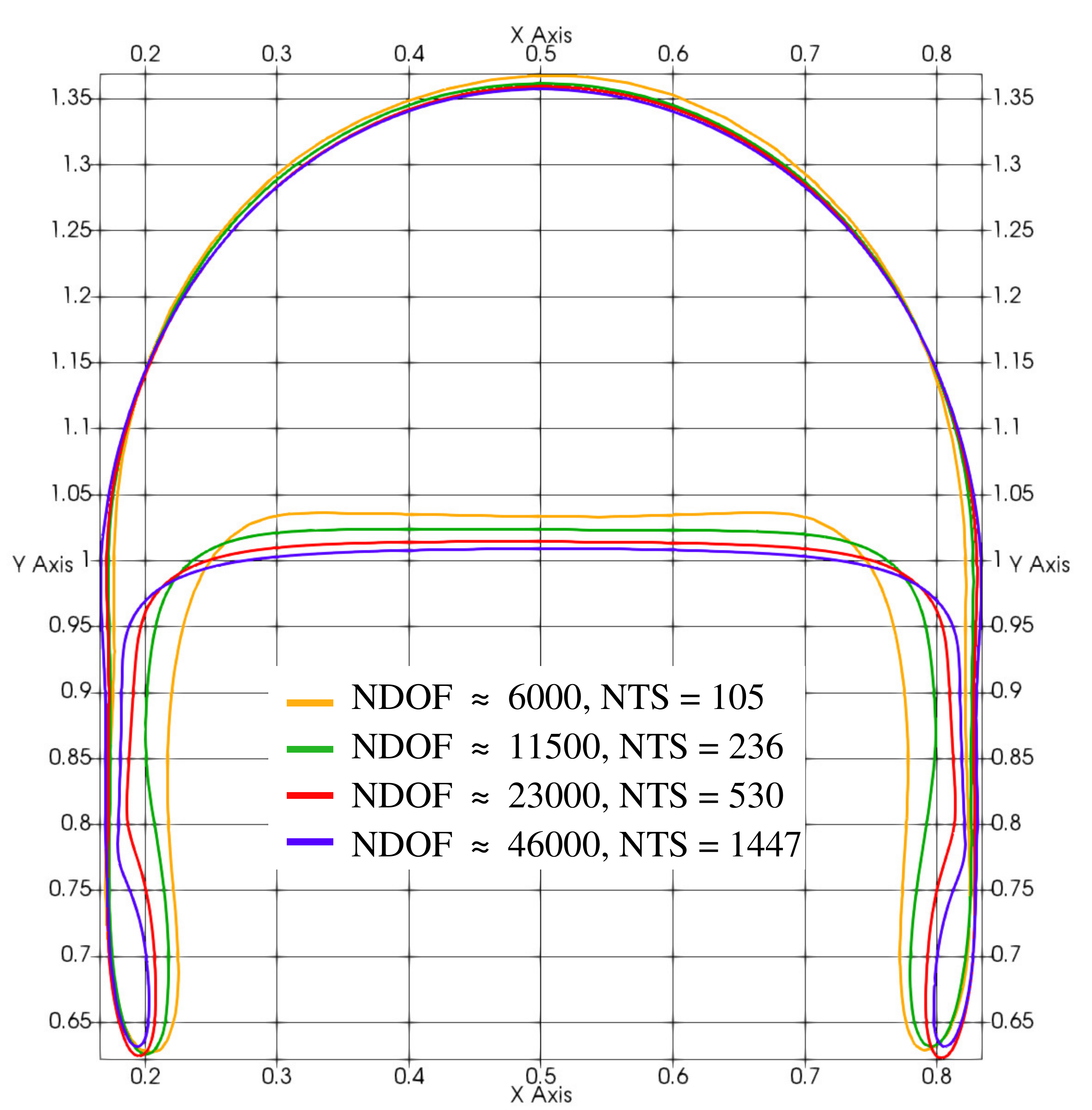}
\caption{Liquid-gas interface for some of the 2D single bubble rise case 2 simulations with adaptive mesh at $t=3$.}
\label{fig:bubble_rise_2D_case_2_shape}
\end{figure}

In case 2, thin bubble filaments develop at lateral extremities of the bubble and distort significantly, while the main part of the bubble remains nearly elliptical with a flat bottom surface. One of the solutions obtained using an adaptive mesh is shown in Fig. \ref{fig:bubble_rise_2D_case_2_result}. The great advantage of anisotropic mesh adaption is demonstrated once again, because both thin filaments are captured and tracked by the adaptive mesh, as shown in Fig. \ref{fig:bubble_rise_2D_case_2_result}(f).\\
A closer look at the final shapes of the bubble in some simulations is presented in Fig. \ref{fig:bubble_rise_2D_case_2_shape}. Convergence is more difficult for this case as the filaments become thinner and thinner with a smaller mesh size and a smaller $\epsilon$. This change of thickness of the bubble filaments with finer meshes affects slightly the whole shape of the bubble. For the simulation with the highest NDOF and NTS, the lateral extremities of the bubble seem to transform into new bubbles which could potentially emerge and breakup.\\

\begin{figure}[htbp]
	\centering
	\begin{subfigure}{0.49\textwidth}
	    \includegraphics[width=\textwidth,trim={20 0 35 35},clip]{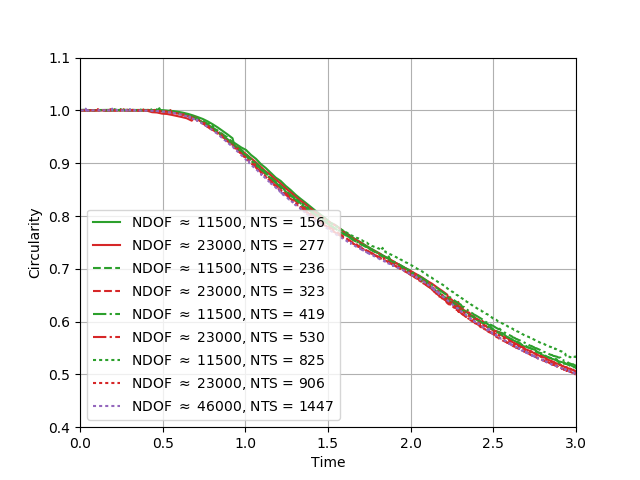}
		\caption{}
	\end{subfigure}%
	~
	\begin{subfigure}{0.49\textwidth}
	    \includegraphics[width=\textwidth,trim={10 0 35 35},clip]{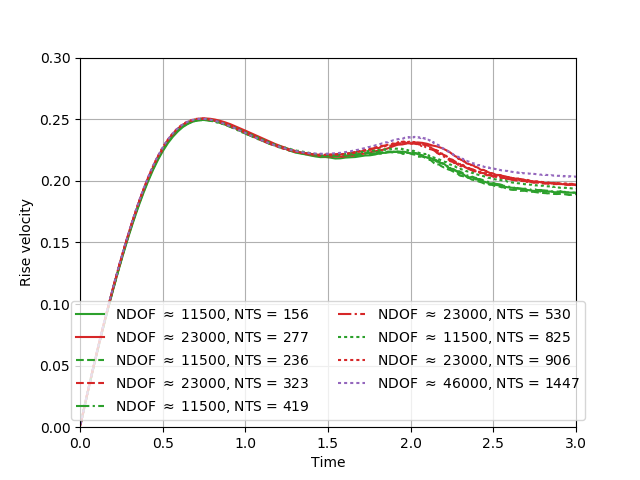}
		\caption{}
	\end{subfigure}%
\caption{Circularity (a) and rise velocity (b) evolution for the 2D single bubble rise case 2 simulations using an adaptive mesh with various NDOF and NTS.}
\label{fig:bubble_rise_2D_case_2_curves}
\end{figure}

Some codes assessed in the benchmark predicted a breakup of the thin filaments \cite{Hysing2009}. This could be clearly observed by looking at an additional parameter of interest, \textit{viz.} the bubble circularity $S_b$ which is defined as
$$S_{b}(t) = \frac{2\pi}{V_b(t)}\sqrt{\frac{\int_\Omega \delta(-\phi(\mathbf{x},t)) \mathrm{d}\Omega}{\pi}}.$$
After the point of breakup, some codes did not show clear convergence on circularity curves and significant differences were found when comparing curves computed using different codes. We provide an explanation of those differences in the following.\\
The TP2D code used the LS method with bilinear FEs for the LS function and a second order time stepping scheme with a weak coupling between Navier-Stokes equations and the LS advection equation. The FreeLIFE code also used the LS method with a second order time stepping scheme and a weak coupling, but with linear FEs for the LS function. An additional algorithm was implemented to ensure global mass conservation. The MooNMD code used linear FEs with conformal meshing of the interface and a second order time stepping scheme with a weak coupling. An Arbitrary Lagrangian-Eulerian (ALE) method was implemented to achieve local mass conservation better. See Ref. \cite{Hysing2009} for more references on each code and how they were used in the benchmark.\\
Based on these facts, the code that could theoretically achieve the best performance regarding local mass conservation was the MooNMD code owing to its ALE method. It is also the only code that could reach a clear convergence on circularity curves for the numerical parameters used in the benchmark. Our method can achieve a similar performance by virtue of the use of a P2 LS function and anisotropic mesh adaption. This is demonstrated by looking at the errors of $V_b$ in Tab. \ref{tab:bubble-rise-2D-case-2-errors} and for case 1 in Tab. \ref{tab:bubble-rise-2D-case-1-errors}.\\
Neither the MooNMD code nor our method predict breakup of the filaments. Moreover, as shown in Fig. \ref{fig:bubble_rise_2D_case_2_curves}(a), the circularity evolution predicted by our method converges at $t=3$ to a value a little greater than $0.5$ which is similar to that predicted by the MooNMD code \cite{Hysing2009}. Our results are also in agreement with other studies where no breakup of the filaments was observed for this benchmark case \cite{Aland2012,Doyeux2013,Wang2015}.\\
The accuracy of the velocity field predicted by our method could still be improved, as shown in Fig. \ref{fig:bubble_rise_2D_case_2_curves}(b). The rise velocity converges to a final value above $0.2$, which matches with the results found in the benchmark for all codes. The convergence is nevertheless not as fast as for the circularity, as shown in Tab. \ref{tab:bubble-rise-2D-case-2-errors} where the results with optimal convergence rate are highlighted in bold. We do not obtain any convergence rate on the square root of NDOF which is close to three for the rise velocity. A multi-criterion error estimator for anisotropic P2 mesh adaption which would take into account the velocity field should hence be considered in future developments.\\
We note however that we used NTSs which were lower than that used by the MooNMD code (between 3,000 and 6,000). This is a limitation of the ALE approach, which is avoided with the Eulerian method used herein with anisotropic mesh adaption. Owing to the P2/P2/P2 scheme, we can reach a high NDOF with a relatively low NEL compared to that used by all codes assessed in the benchmark. This is of particular interest for our approach regarding mesh adaption, because the computational cost of mesh adaption is related to NEL (Subsec. \ref{subsec:method:adapt}).\\

\begin{table}[htbp]
    \centering
    \begin{tabular}{r||c|c|c|c}
        $\epsilon$ & $0.05$ & $0.025$ & $0.0125$ & $0.00625$ \\
        NEL & $\approx$ 740 & $\approx$ 1,400 & $\approx$ 2,800 & $\approx$ 5,600 \\
        NDOF & $\approx$ 6,000 & $\approx$ 11,500 & $\approx$ 23,000 & $\approx$ 46,000 \\ \hline
        NTS & 105 & 156 &  &  \\
        Error($V_b$) & $4.38 \times 10^{-2}$ & $4.31 \times 10^{-2}$ &  &  \\
        Error($S_b$) & $2.38 \times 10^{-2}$ & $1.39 \times 10^{-2}$ &  & \\
        Error($y_b$) & $1.02 \times 10^{-2}$ & $\mathbf{3.61 \times 10^{-3}}$ &  & \\
        Error($v_b$) & $8.70 \times 10^{-2}$ & $\mathbf{3.84 \times 10^{-2}}$ &  & \\ \hline
        NTS & 203 & 236 & 323 & \\
        Error($V_b$) & $1.54 \times 10^{-2}$ & $1.83 \times 10^{-2}$ & $2.10 \times 10^{-2}$ & \\
        Error($S_b$) & $2.27 \times 10^{-2}$ & $\mathbf{9.59 \times 10^{-3}}$ & $4.73 \times 10^{-3}$ & \\
        Error($y_b$) & $1.13 \times 10^{-2}$ & $5.14 \times 10^{-3}$ & $\mathbf{1.52 \times 10^{-3}}$ & \\
        Error($v_b$) & $8.05 \times 10^{-2}$ & $4.12 \times 10^{-2}$ & $\mathbf{1.65 \times 10^{-2}}$ & \\ \hline
        NTS & 403 & 419 & 530 & \\
        Error($V_b$) & $3.13 \times 10^{-3}$ & $2.70 \times 10^{-3}$ & $6.79 \times 10^{-3}$ & \\
        Error($S_b$) & $2.00 \times 10^{-2}$ & $1.21 \times 10^{-2}$ & $\mathbf{2.77 \times 10^{-3}}$ & \\
        Error($y_b$) & $1.46 \times 10^{-2}$ & $\mathbf{5.45 \times 10^{-3}}$ & $2.37 \times 10^{-3}$ & \\
        Error($v_b$) & $8.34 \times 10^{-2}$ & $\mathbf{3.64 \times 10^{-2}}$ & $1.76 \times 10^{-2}$ & \\ \hline
        NTS & 804 & 825 & 906 & 1447 \\
        Error($V_b$) & $1.86 \times 10^{-2}$ & $8.34 \times 10^{-3}$ & $\mathbf{1.64 \times 10^{-3}}$ & $1.19 \times 10^{-3}$ \\
        Error($S_b$) & $2.31 \times 10^{-2}$ & $2.25 \times 10^{-2}$ & $\mathbf{3.39 \times 10^{-3}}$ & \\
        Error($y_b$) & $1.39 \times 10^{-2}$ & $\mathbf{4.39 \times 10^{-3}}$ & $2.67 \times 10^{-3}$ & \\
        Error($v_b$) & $7.21 \times 10^{-2}$ & $\mathbf{2.75 \times 10^{-2}}$ & $1.56 \times 10^{-2}$ & \\ \hline
    \end{tabular}
    \caption{Relative $L^2$ errors for 2D single bubble rise case 2 simulations using different space (NDOF) and time (NTS) discretizations, with anisotropic mesh adaption. All results with convergence rate with respect to the square root of NDOF higher than 2.5 are highlighted in bold.}
    \label{tab:bubble-rise-2D-case-2-errors}
\end{table}

\REV{Following the diagonal in Tab. \ref{tab:bubble-rise-2D-case-2-errors}, the total computation times on a machine with a 20-core \textit{Intel Xeon E5-2630 v4 @ 2.20GHz} CPU and a 3072-core \textit{NVIDIA GTX TITAN X} GPU were about 10min, 28min, 1h28min and 4h32min, respectively. For all case 1 and case 2 simulations, the} computation time was mainly spent on the Newton-Raphson solution of the RBVMS formulation whereas the computational cost of LS reinitialization was negligible (smaller than $1\%$). For the simulations with an adaptive mesh, the computation time of error estimation, SPR, metric computation, mesh adaption and fields transfer represented between $1$ and $15\%$ of the total computation time. In addition, we did not observe any significant mass loss due to P2 fields transfer after mesh adaption. Instead, we observed an improved mass conservation at equal NDOF with P2 mesh adaption. Therefore, we can conclude that there is a great merit in using an adaptive P2 mesh.

\subsection{Single bubble rise (3D)}

\begin{figure}[htbp]
	\centering
	\begin{subfigure}{0.4\textwidth}
	    \includegraphics[height=9cm,trim={30mm 10mm 0mm 15mm},clip]{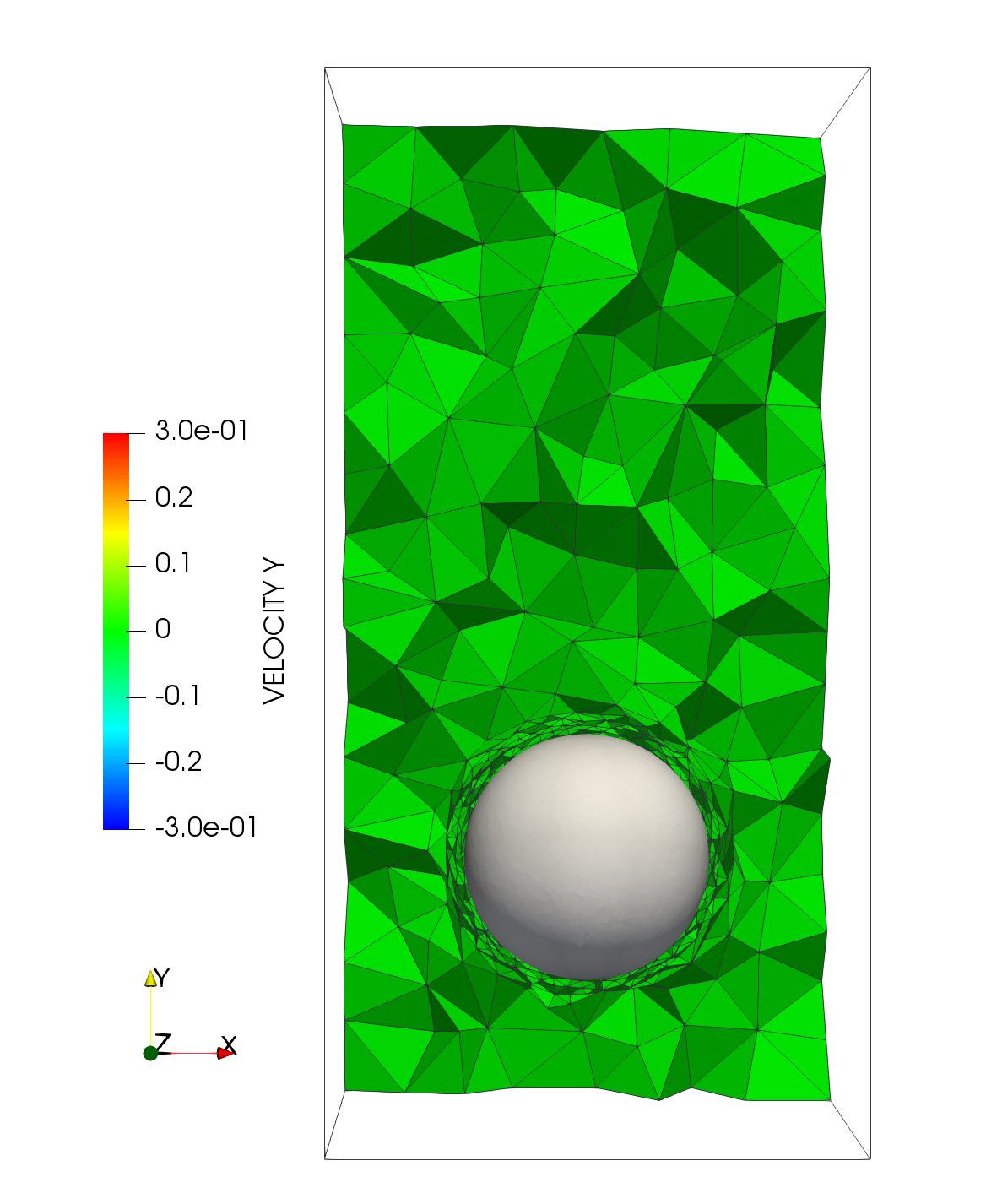}
		\caption{}
	\end{subfigure}%
	~
	\begin{subfigure}{0.3\textwidth}
	    \includegraphics[height=9cm,trim={120mm 10mm 40mm 15mm},clip]{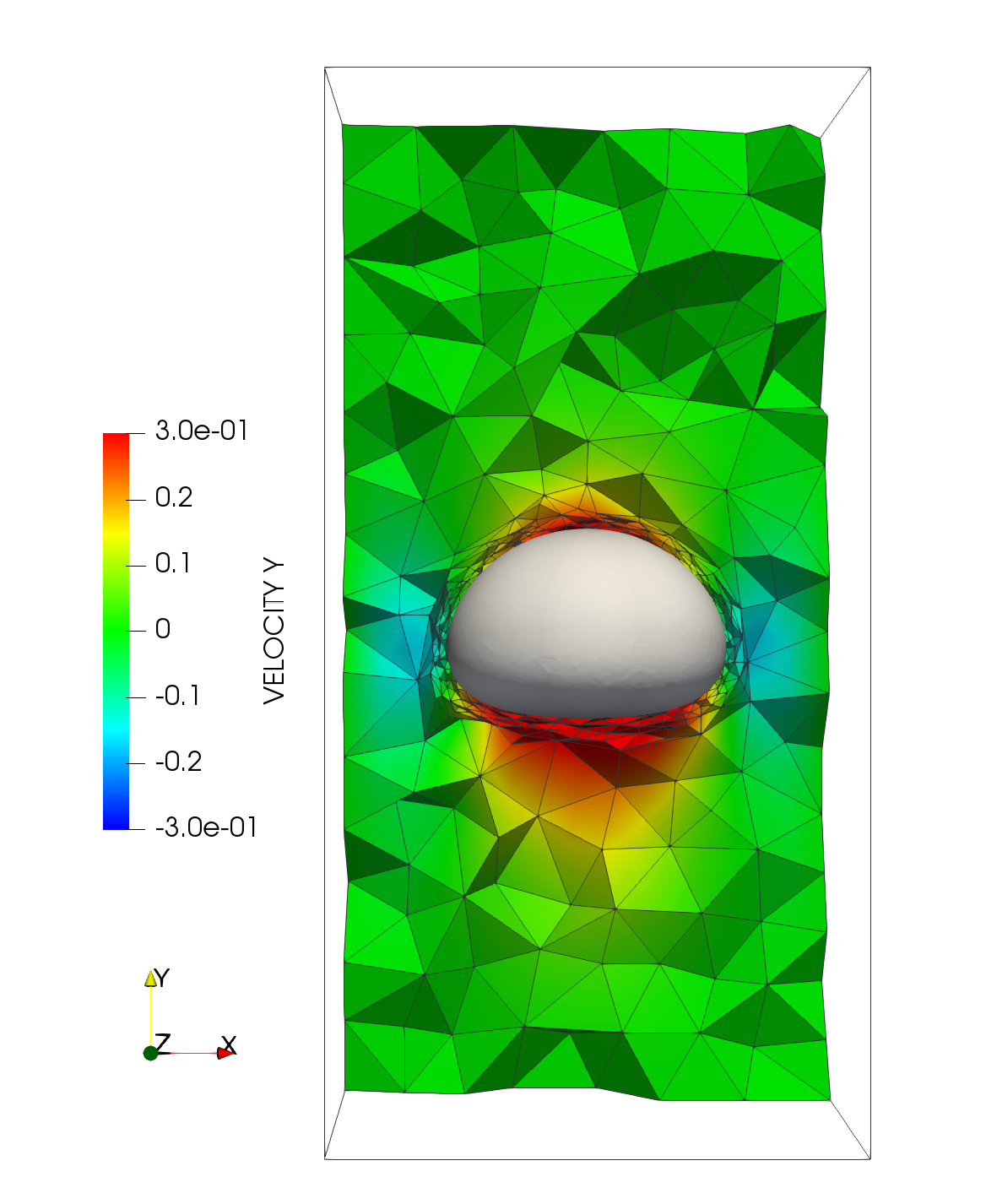}
		\caption{}
	\end{subfigure}%
	~
	\begin{subfigure}{0.3\textwidth}
	    \includegraphics[height=9cm,trim={120mm 10mm 40mm 15mm},clip]{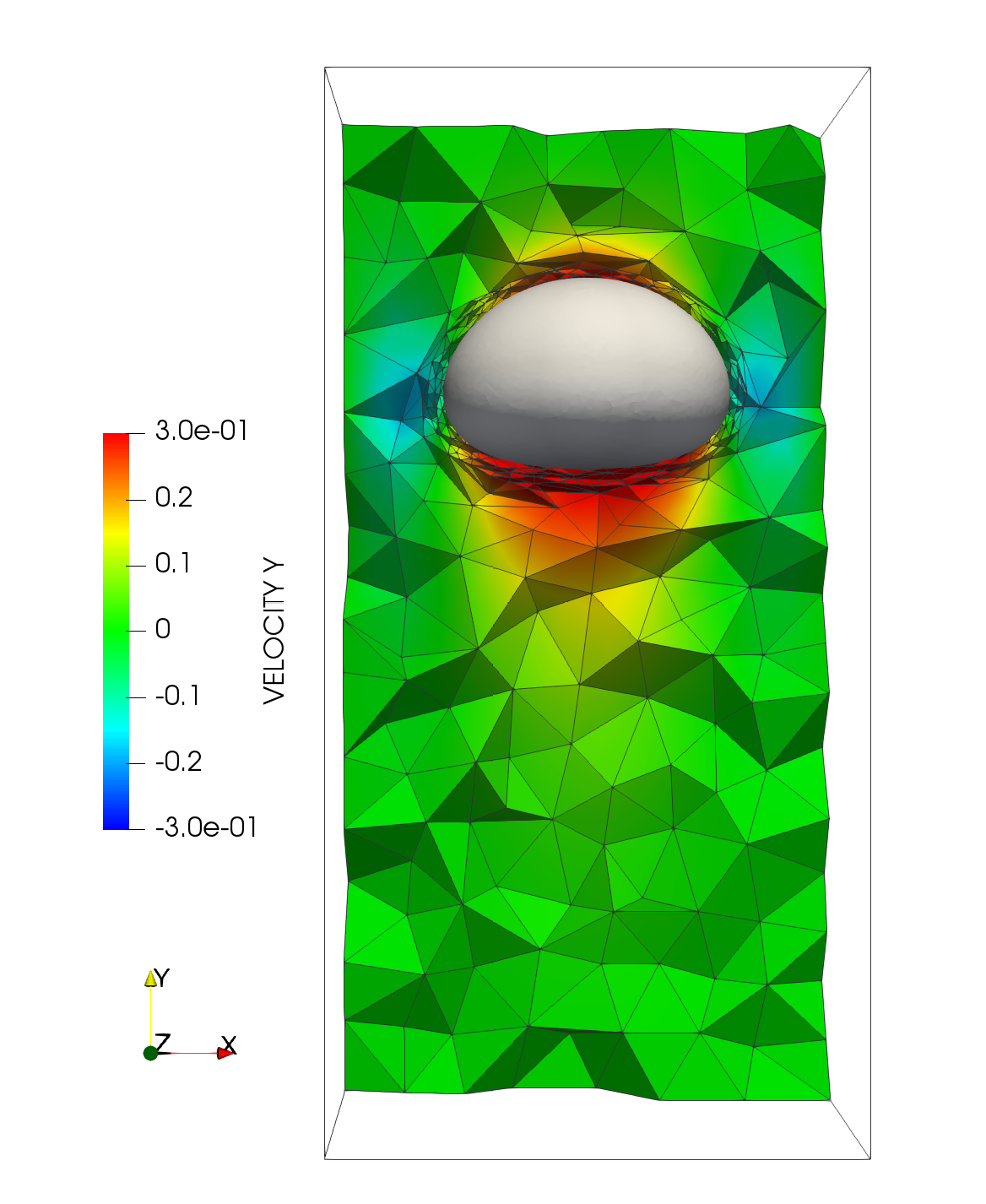}
		\caption{}
	\end{subfigure}%
\caption{Velocity field in rise direction $\mathbf{v}.\mathbf{e}_y$ for the 3D single bubble rise case 1 simulations using an adaptive mesh with NDOF $\approx 375000$ and NTS $\approx 200$ at: (a) $t = 0$, (b) $t = 1.5$, (c) $t=3.0$.}
\label{fig:bubble_rise_3D_case_1_result}
\end{figure}

\begin{figure}[htbp]
	\centering
	\begin{subfigure}{0.4\textwidth}
	    \includegraphics[height=9cm,trim={30mm 10mm 10mm 15mm},clip]{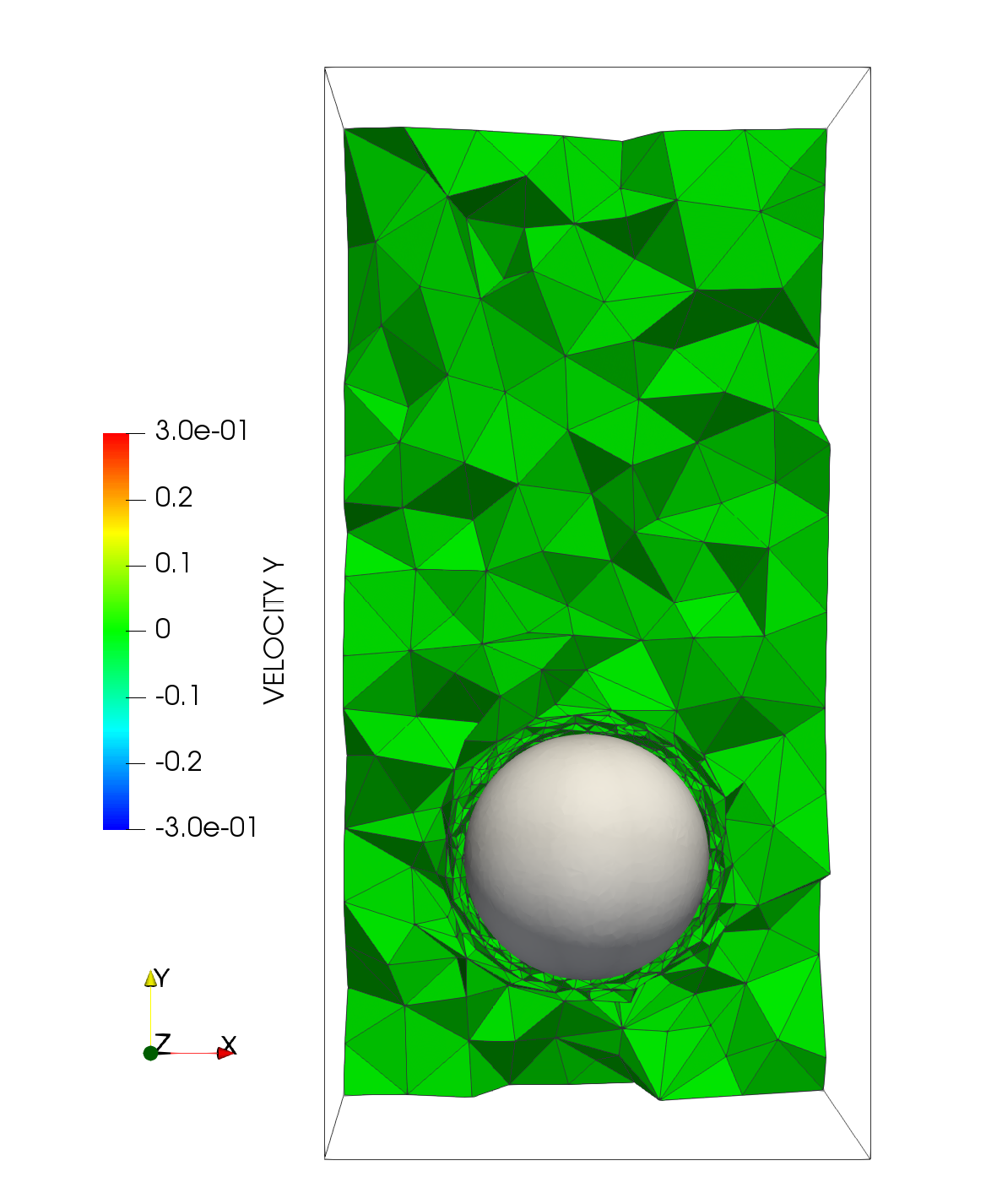}
		\caption{}
	\end{subfigure}%
	~
	\begin{subfigure}{0.3\textwidth}
	    \includegraphics[height=9cm,trim={120mm 10mm 40mm 15mm},clip]{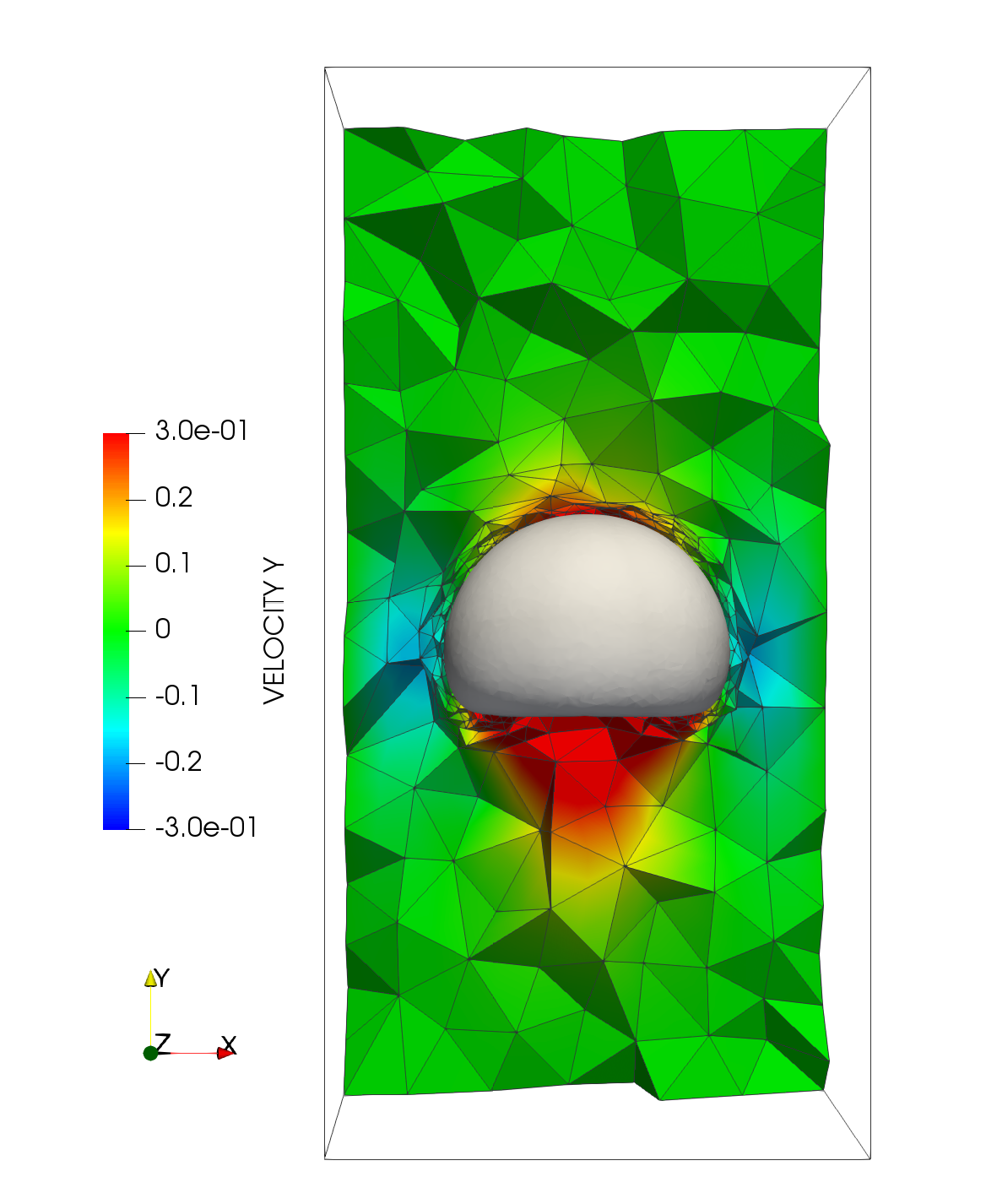}
		\caption{}
	\end{subfigure}%
	~
	\begin{subfigure}{0.3\textwidth}
	    \includegraphics[height=9cm,trim={120mm 10mm 40mm 15mm},clip]{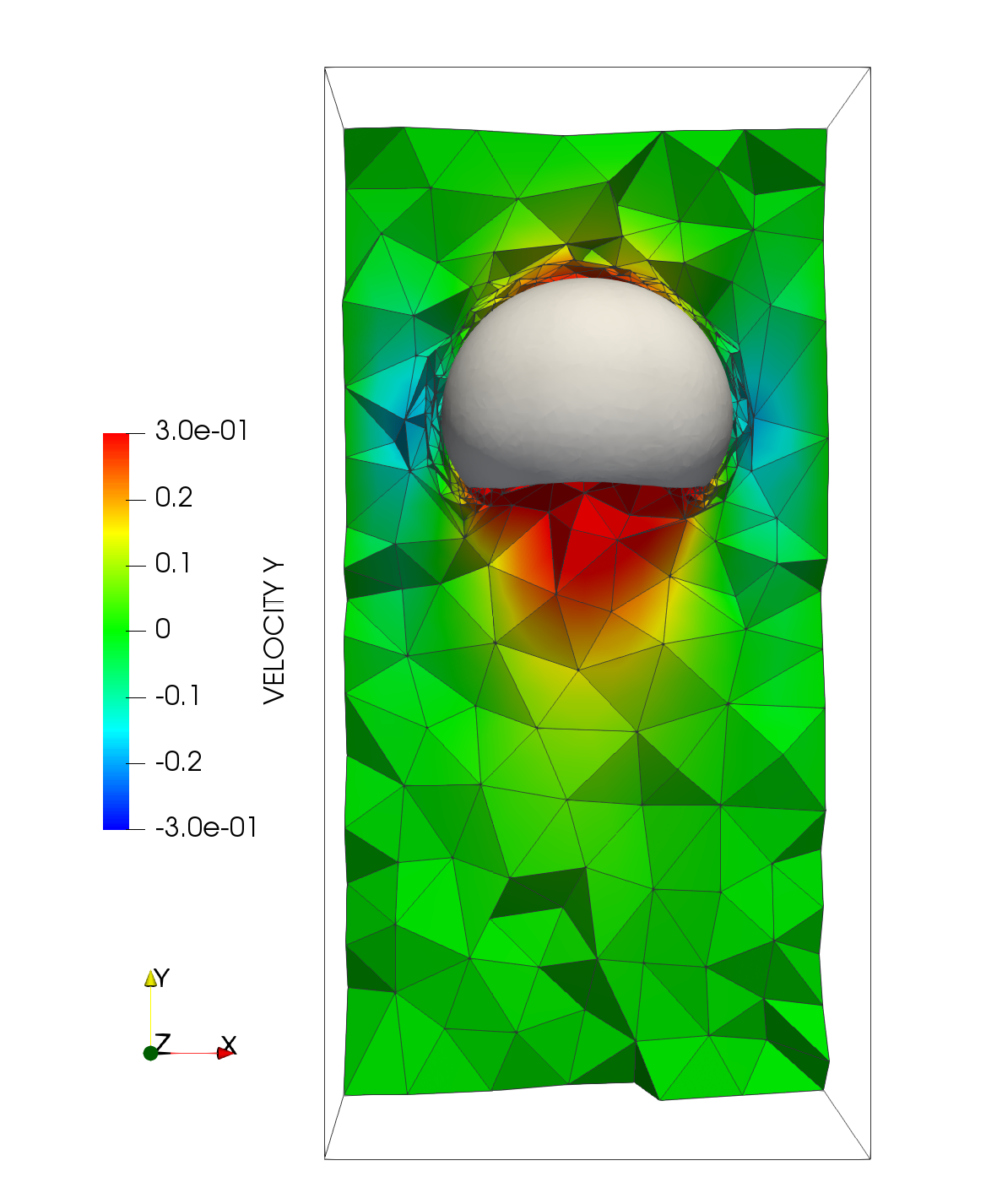}
		\caption{}
	\end{subfigure}%
\caption{Velocity field in rise direction for the 3D single bubble rise case 2 simulations using an adaptive mesh with NDOF $\approx 375000$ and NTS $\approx 200$ at: (a) $t = 0$, (b) $t = 1.5$, (c) $t=3.0$.}
\label{fig:bubble_rise_3D_case_2_result}
\end{figure}

\begin{figure}[htbp]
	\centering
	\begin{subfigure}{0.3\textwidth}
	    \includegraphics[width=\textwidth,trim={0 60mm 0 15mm},clip]{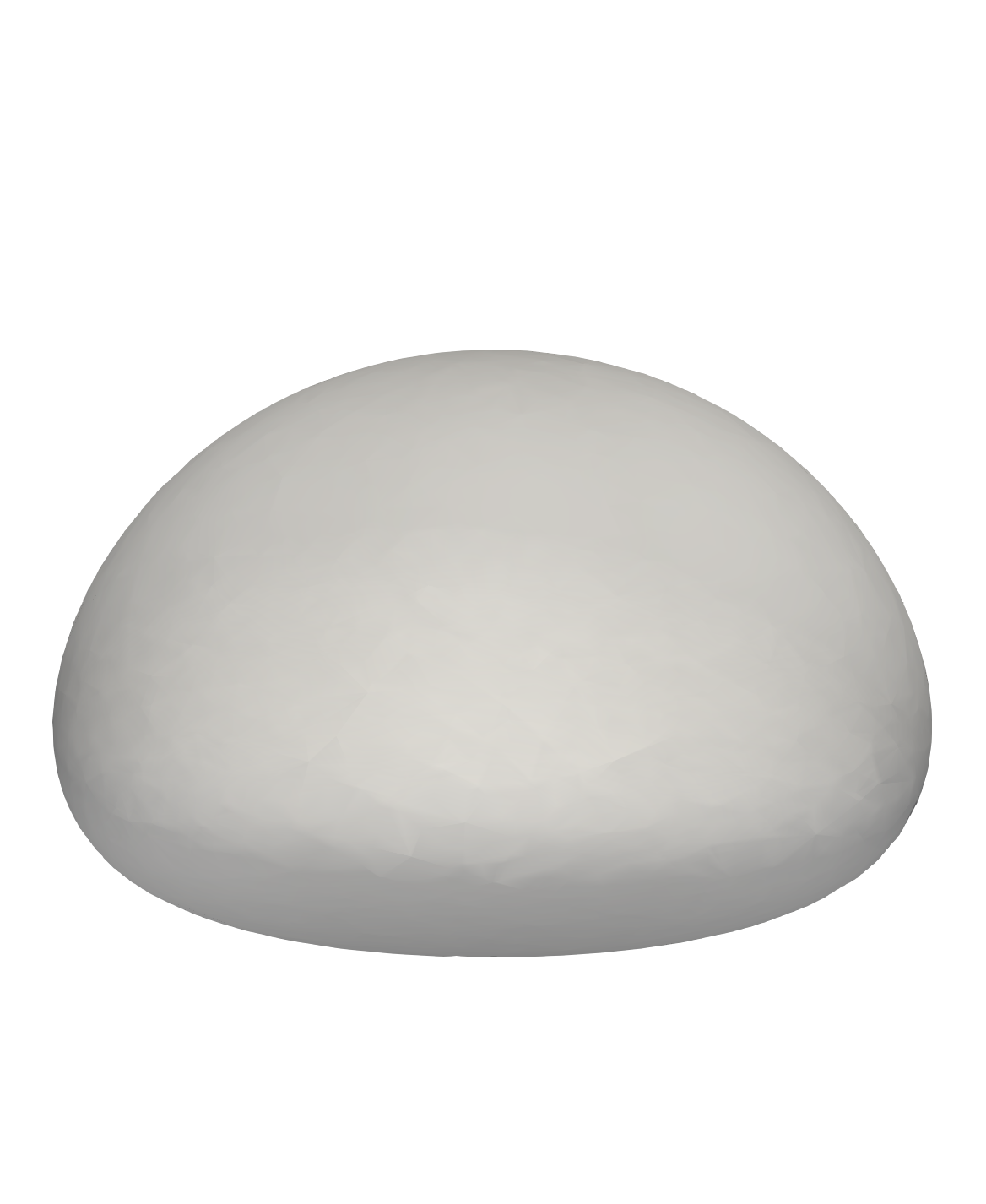}
		\caption{}
	\end{subfigure}%
	~
	\begin{subfigure}{0.3\textwidth}
	    \includegraphics[width=\textwidth,trim={0 60mm 0 15mm},clip]{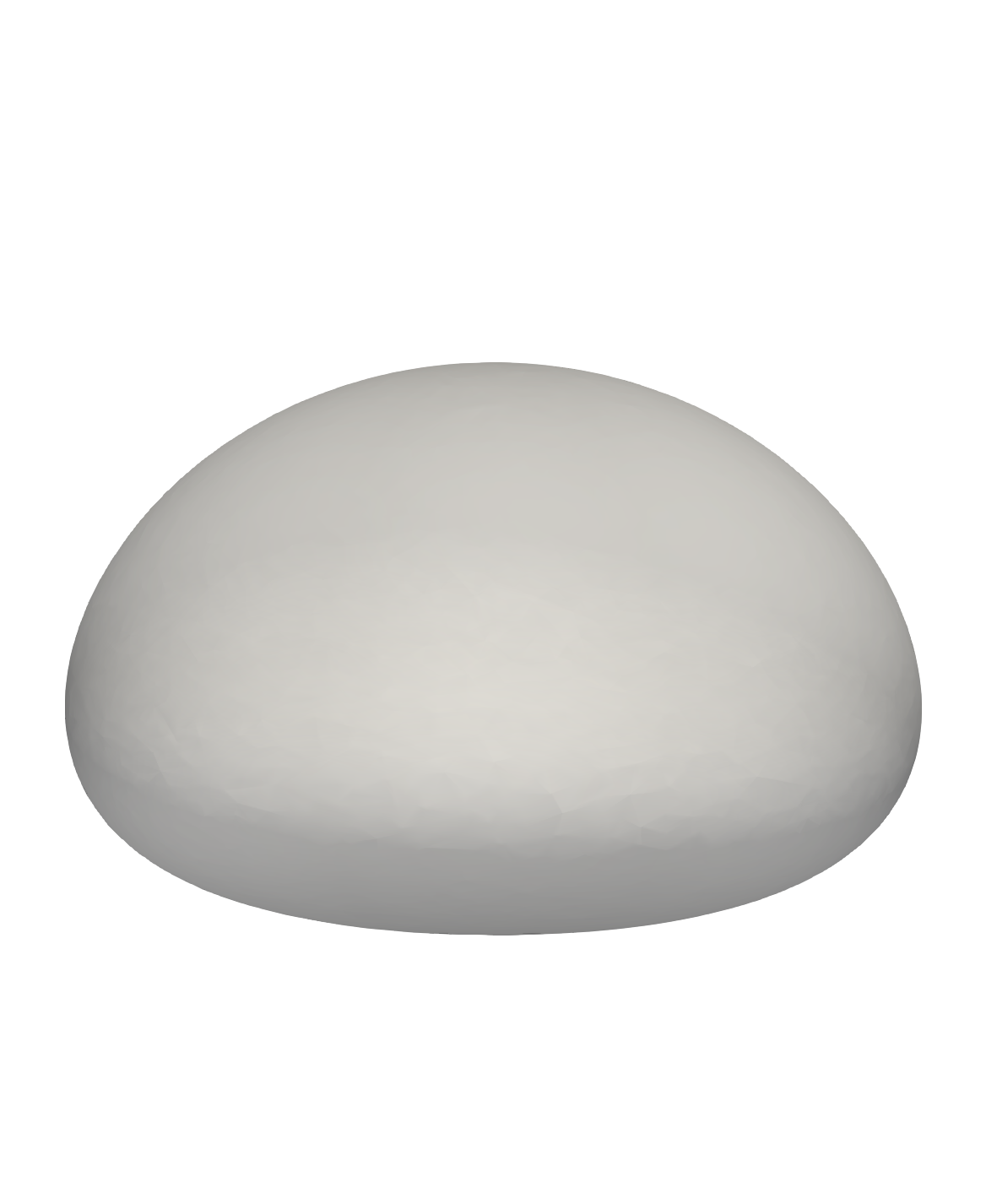}
		\caption{}
	\end{subfigure}%
	~
	\begin{subfigure}{0.366267123\textwidth}
	    \includegraphics[width=\textwidth,trim={0 60mm 0 15mm},clip]{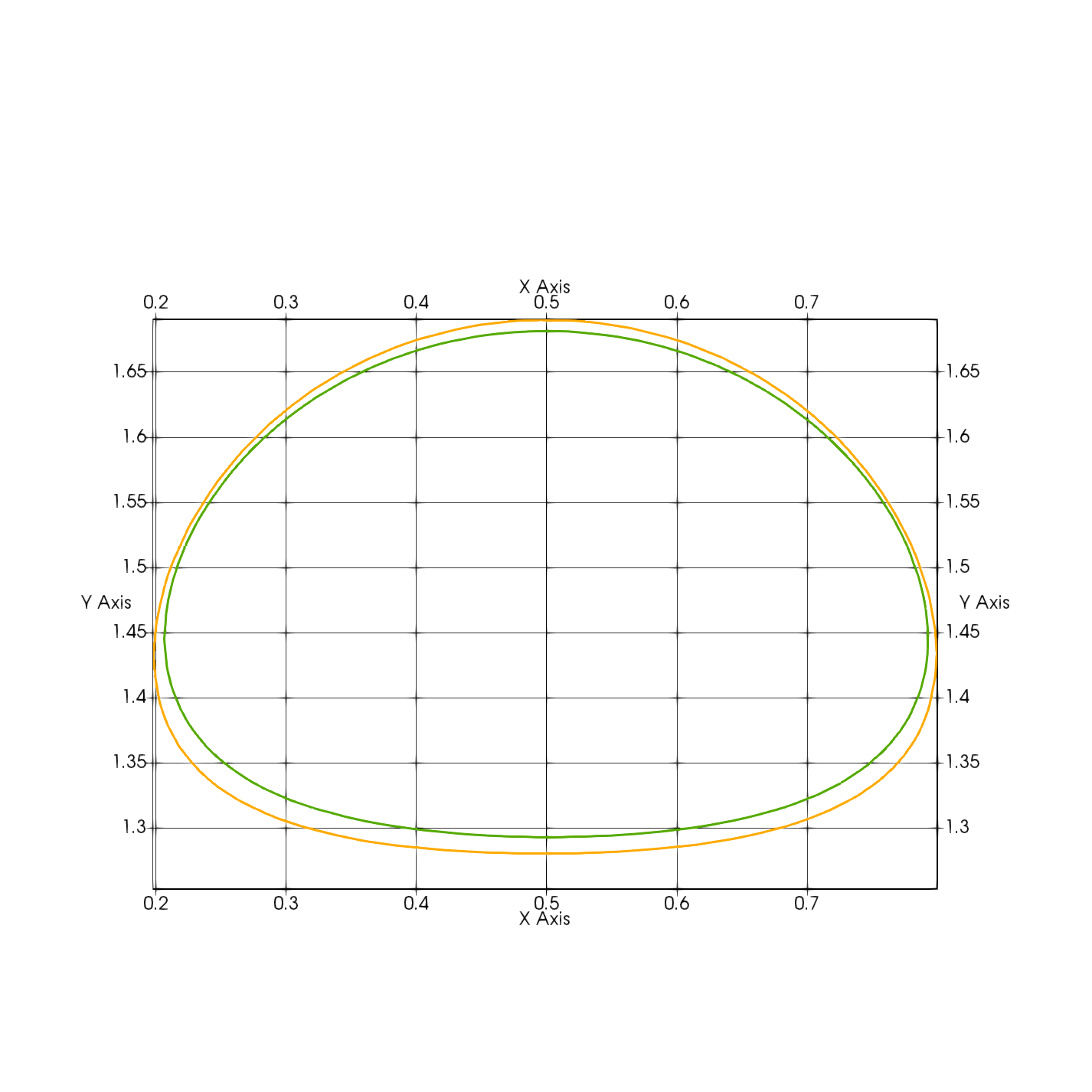}
		\caption{}
	\end{subfigure}%
	\hfill
	\begin{subfigure}{0.3\textwidth}
	    \includegraphics[width=\textwidth,trim={0 60mm 0 15mm},clip]{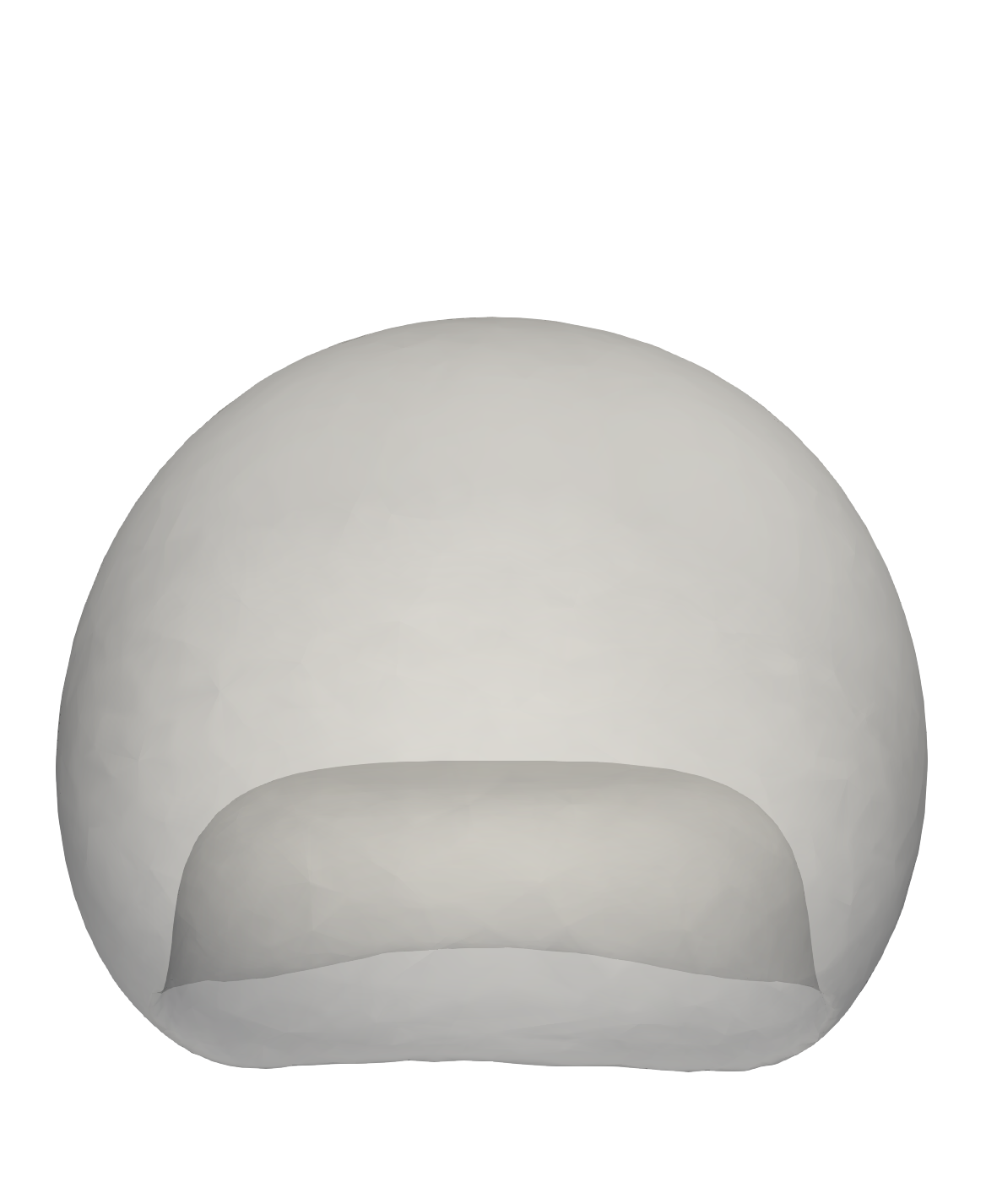}
		\caption{}
	\end{subfigure}%
	~
	\begin{subfigure}{0.3\textwidth}
	    \includegraphics[width=\textwidth,trim={0 60mm 0 15mm},clip]{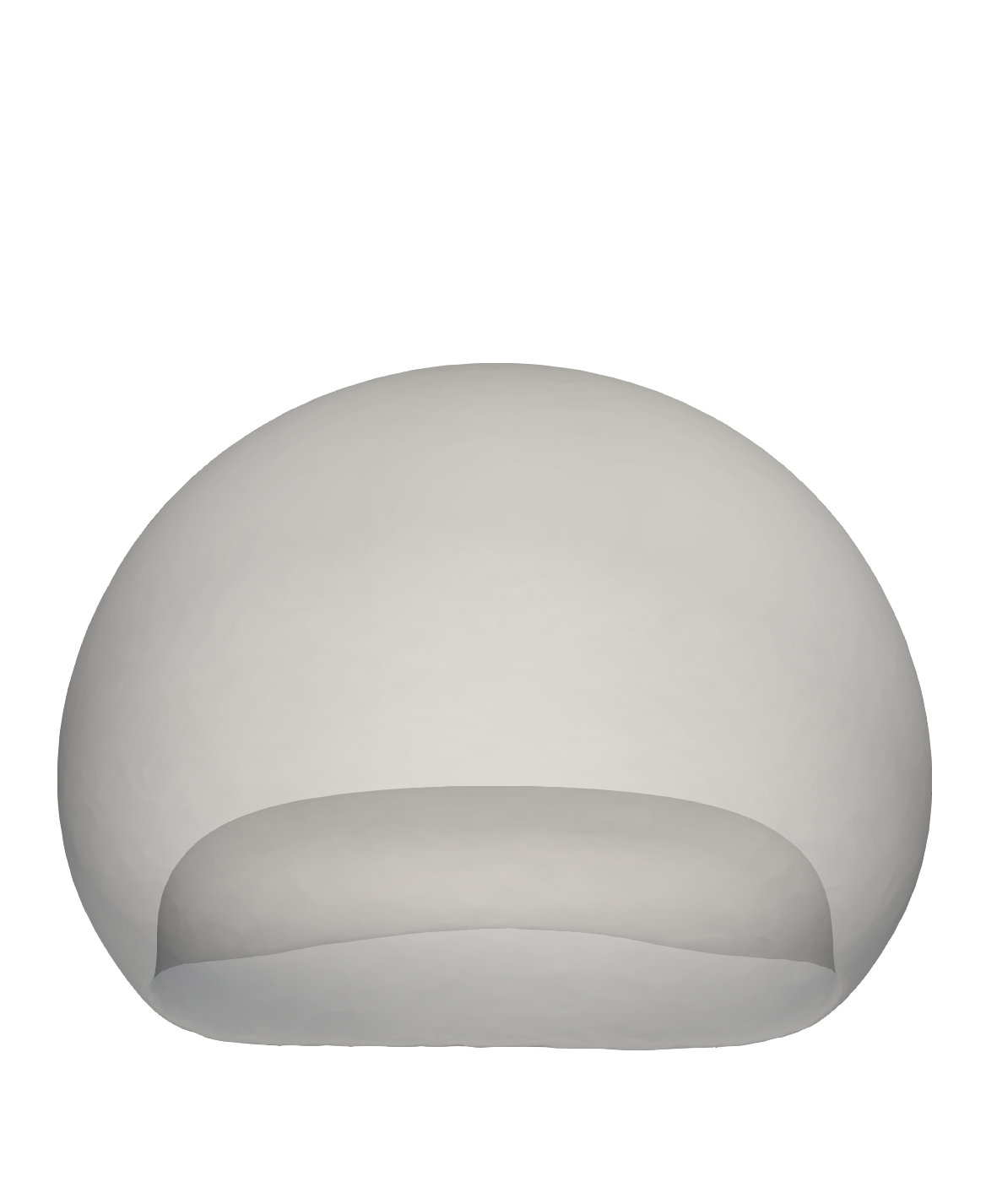}
		\caption{}
	\end{subfigure}%
	~
	\begin{subfigure}{0.366267123\textwidth}
	    \includegraphics[width=\textwidth,trim={0 60mm 0 15mm},clip]{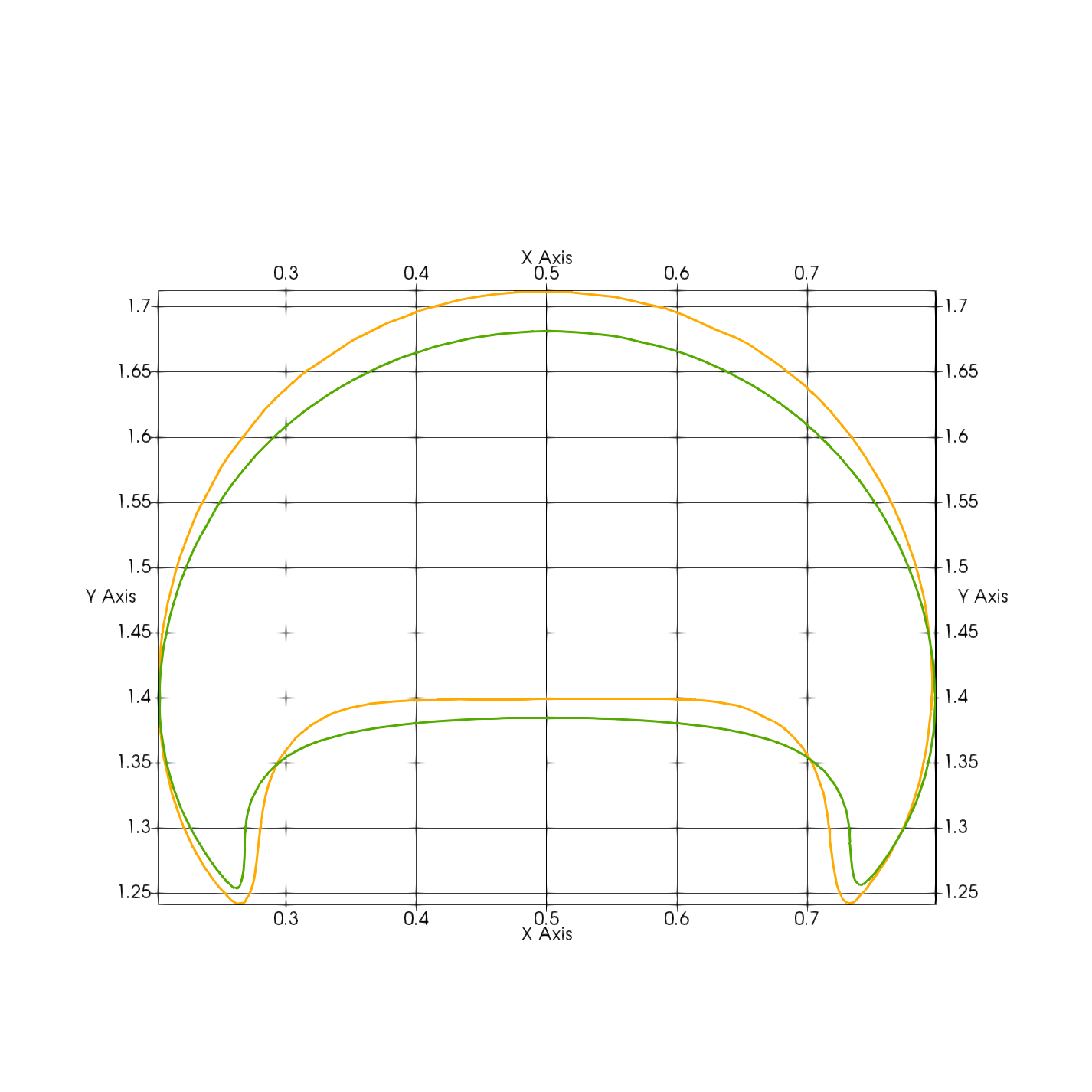}
		\caption{}
	\end{subfigure}%
\caption{Shapes with transparent interface at $t=3$ for the 3D single bubble rise simulations: (a) case 1 using the coarse discretization, (b) case 1 using the fine discretization, (d) case 2 using the coarse discretization, (e) case 2 using the fine discretization. Midsection of the interface using the coarse (orange) and fine (green) discretizations: (c) superposition of (a) and (b) for case 1, (f) superposition of (d) and (e) for case 2.}
\label{fig:bubble_rise_3D_shapes}
\end{figure}

Different 3D versions of the single bubble rise benchmark have been proposed in the literature \cite{Safi2017,Karyofylli2018,Metivet2018,Lin2018}. Here we opt for a simple 3D extension where the computation domain is a box of dimensions $1 \times 2 \times 1$. A spherical bubble of radius $R = \REV{0.25}$ is initially placed at $(0.5,0.5,0.5)$. No-slip boundary conditions are applied at all boundaries. Fluid and flow properties with two cases are identical to those of the 2D version. This corresponds to the 3D setup used in Refs. \cite{Safi2017,Lin2018}. Note that 3D results are not expected to be comparable to 2D results due to the change of boundary conditions and the extra dimension.\\
Two discretizations are used for each case. The coarse discretization has an adaptive mesh of NEL $\approx 25,000$ and NDOF $\approx 175,000$ for a NTS $\approx 100$, while the fine discretization has respectively NEL $\approx 56,000$ and NDOF $\approx 375,000$ for a NTS $\approx 200$. The value of $\epsilon$ is $0.05$ for the coarse discretization and $0.025$ for the fine one.\\
The results with the fine discretization are shown for case 1 in Fig. \ref{fig:bubble_rise_3D_case_1_result} and for case 2 in Fig. \ref{fig:bubble_rise_3D_case_2_result}. The final shape of the bubble for each case is similar to that obtained by other authors in the literature \cite{Safi2017,Lin2018}. Contrary to the 2D version, there is no excessive stretching of the bubbles into filaments that might breakup for case 2, which is also in agreement with the literature \cite{Safi2017,Lin2018}.\\
A closer look at the final shapes is shown in Fig. \ref{fig:bubble_rise_3D_shapes}. There is clearly an overestimation of the bubble volume in both cases for the coarse discretization, which can be attributed to the low order of the time stepping scheme based on the 2D simulations. The final shapes are nevertheless smooth, even using the coarse discretization. In comparison with other results for a similar number of elements found in the literature \cite{Safi2017,Lin2018,Yan2018iga}, this result shows the great advantage of using an anisotropic adaptive P2 interpolation for the LS function.\\

\begin{figure}[htbp]
	\centering
	\begin{subfigure}{0.49\textwidth}
	    \includegraphics[width=\textwidth,trim={10 0 35 35},clip]{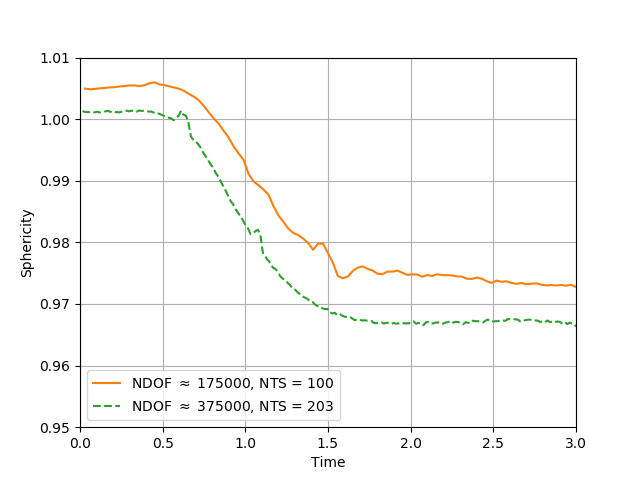}
		\caption{}
	\end{subfigure}%
	~
	\begin{subfigure}{0.49\textwidth}
	    \includegraphics[width=\textwidth,trim={10 0 35 35},clip]{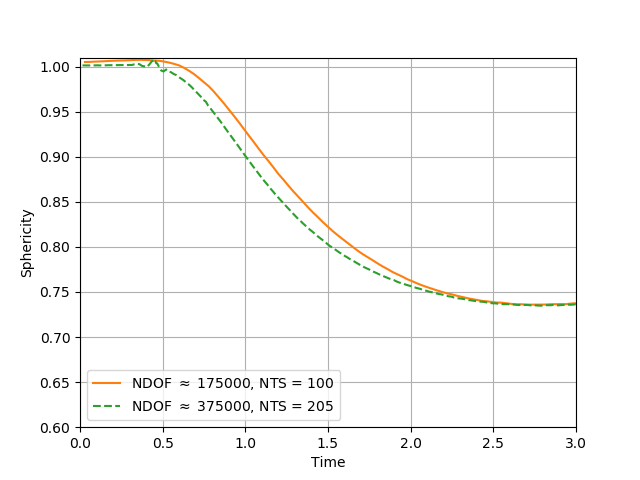}
		\caption{}
	\end{subfigure}%
	\hfill
	\begin{subfigure}{0.49\textwidth}
	    \includegraphics[width=\textwidth,trim={10 0 35 35},clip]{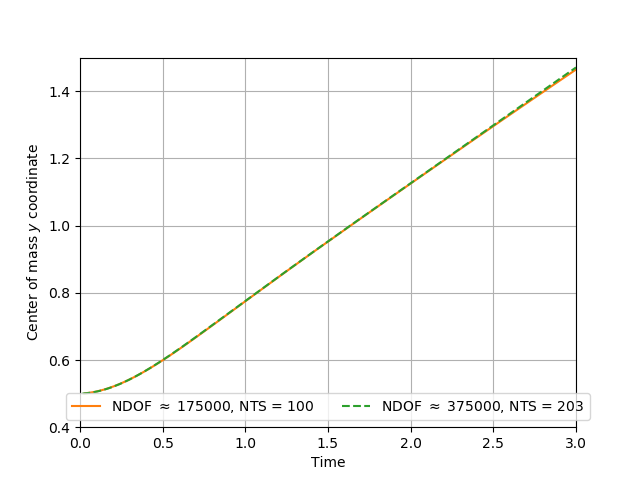}
		\caption{}
	\end{subfigure}%
	~
	\begin{subfigure}{0.49\textwidth}
	    \includegraphics[width=\textwidth,trim={10 0 35 35},clip]{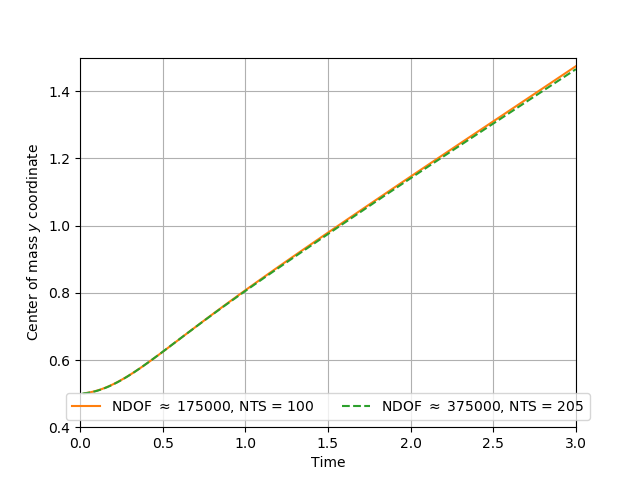}
		\caption{}
	\end{subfigure}%
	\hfill
	\begin{subfigure}{0.49\textwidth}
	    \includegraphics[width=\textwidth,trim={10 0 35 35},clip]{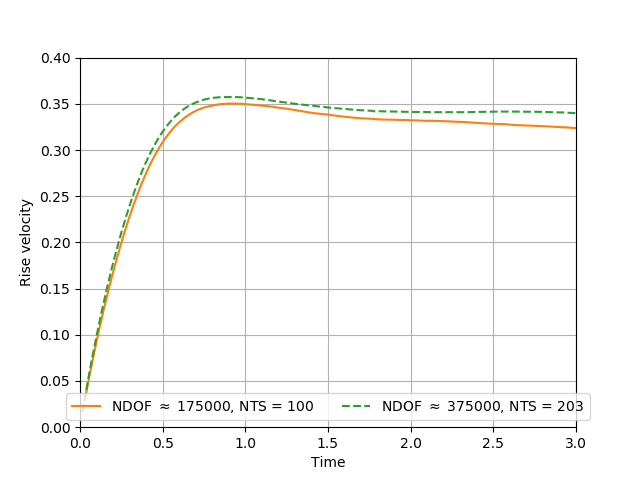}
		\caption{}
	\end{subfigure}%
	~
	\begin{subfigure}{0.49\textwidth}
	    \includegraphics[width=\textwidth,trim={10 0 35 35},clip]{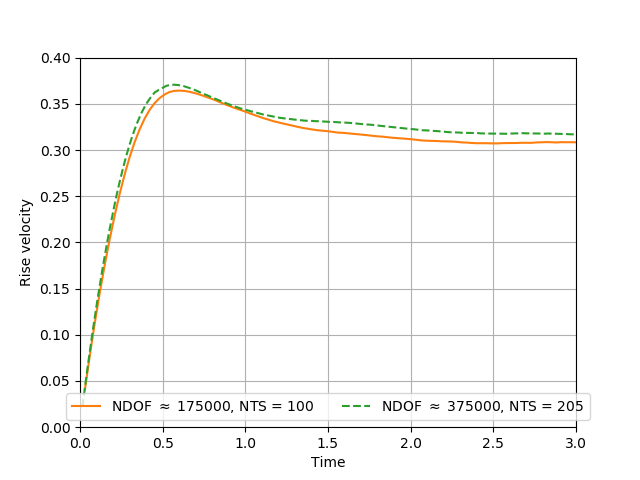}
		\caption{}
	\end{subfigure}%
\caption{3D single bubble rise simulation results using an adaptive mesh with various NDOF and NTS. Case 1: (a) sphericity, (c) center of mass $y$ coordinate, (e) rise velocity. Case 2: (b) sphericity, (d) center of mass $y$ coordinate, (f) rise velocity. Scales may differ from left to right.}
\label{fig:bubble_rise_3D_curves}
\end{figure}

The computed parameters of interest are the same as for the 2D case, except that $V_b(t)$ now represents the bubble volume and the bubble circularity $S_{b}$ is replaced by the bubble sphericity given by:
$$S_{b}(t) = \frac{4\pi}{V_b(t)}\left(\sqrt[3]{\frac{3}{4\pi}\int_\Omega \delta(-\phi(\mathbf{x},t)) \mathrm{d}\Omega}\right)^2.$$
Time evolution of each parameter of interest is reported for both cases in Fig. \ref{fig:bubble_rise_3D_curves}. For case 1, the errors of $V_b$ are $1.32 \times 10^{-1}$ for the coarse discretization and $5.43 \times 10^{-2}$ for the fine discretization, while for case 2 they are respectively $1.38 \times 10^{-1}$ and $5.65 \times 10^{-2}$. As expected, the errors are slightly higher for case 2, while this difference is quite low compared to the differences between the two simulation results shown in Fig. \ref{fig:bubble_rise_3D_shapes}. This proves that our numerical method is robust and can achieve a similar accuracy for very different bubble deformations.\\
The error of $y_b$ is very low for both cases, as shown in Fig. \ref{fig:bubble_rise_3D_curves}(c,d) where the overall position of the bubble is well captured. In order to describe the bubble's shape, a finer discretization is required, as shown by the sphericity curves in Fig. \ref{fig:bubble_rise_3D_curves}(a,b). Small oscillations can be seen for case 1 in Fig. \ref{fig:bubble_rise_3D_curves}(a), which are due to local mass loss during fields transfer after mesh adaption. Similar oscillations can be seen for case 2 in Fig. \ref{fig:bubble_rise_3D_curves}(b). Their amplitude is clearly low, however compared to the overall evolution of the sphericity. The rise velocity is underestimated with the coarse discretization, as shown in Fig. \ref{fig:bubble_rise_3D_curves}(e,f). This was also observed in Ref. \cite{Lin2018}. The differences between the two discretizations are however quite low. Both the instant at which the maximum rise velocity is reached and the value of the maximum rise velocity are well predicted with the coarse discretization.\\

All \REV{3D single bubble rise} simulations were run on a machine with a 20-core \textit{Intel Xeon E5-2630 v4 @ 2.20GHz} CPU, and a 3072-core \textit{NVIDIA GTX TITAN X} GPU. The total computation time was about 9h for coarse discretization simulations and 34h for fine discretization simulations. About $65\%$ of this computation time was spent on Newton-Raphson solution, $7\%$ on the mesh adaption algorithm, $23\%$ on fields transfer after mesh adaption and $5\%$ on LS reinitialization. The computation time of the other operations such as error estimation and the SPR operation was negligible.\\
The computational cost caused by the modification of the mesh at each time increment was clearly more significant for these 3D simulations. However, this was mainly due to the fields transfer operation and in particular to the search of the containing element of the old mesh for each node of the new mesh. This could be significantly reduced by the implementation of a parallel version of this search on the GPU, as we did for LS reinitialization.\\
The increase in the computational cost by anisotropic P2 mesh adaption is marginal and acceptable given the drastic improvements of accuracy in the 2D simulations.

\section{Conclusions}

Incompressible two phase flows with surface tension were modeled through the Navier Stokes equations, the Level-Set (LS) method and the Continuum Surface Force (CSF) model. The equal-order piece-wise quadratic (P2) finite element discretization was used for the simultaneous solution of the velocity, the pressure and the LS function. The Navier-Stokes equations with the CSF model and the LS advection equation were strongly coupled in a residual-based variational multiscale formulation and solved using a Newton-Raphson scheme with numerical differentiation. The fully implicit formulation was shown to be robust and stable even for large time steps.\\
A balanced-force implementation was used for the CSF model. This implementation could be achieved in a straightforward manner with no need to rely on the numerically complex projections of pressure gradient, because the same discretization was used for the pressure field and the regularized Heaviside function. This implementation was shown to reduce errors of both the velocity and pressure fields below the tolerance used for the Newton-Raphson solver on a static problem using exact curvature.\\
The original approach based on anisotropic P2 mesh adaption was proposed to automatically refine, stretch and orient elements along the interface. As the interface was represented by the zero isolevel of an LS function, the new error estimator targeting the P2 interpolation error of the regularized Heaviside function was developed to define a directional error metric field. The continuous mesh framework was used to convert local directional errors into local directional mesh sizes with a control of the global complexity. This strategy was shown to be more accurate than a uniform fixed mesh for a same number of degrees of freedom, and to enable reductions of the number of degrees of freedom with a factor up to 15 for a same accuracy. In addition, the convergence rates and the mass conservation properties could be improved by virtue of P2 interpolation.

\section*{Acknowledgments}

The authors would like to acknowledge the financial support of the French Agence Nationale de la Recherche to COMP3DRE project (grant number: Projet-ANR-16-CE08-0042).

\bibliography{manuscript}

\end{document}